\documentclass[12pt]{article}

\usepackage[utf8]{inputenc}
\usepackage{amsmath,amssymb,amsthm}
\usepackage{bm,mathrsfs}
\usepackage{graphicx}
\usepackage{booktabs}
\usepackage{flafter}
\usepackage{placeins,color}
\usepackage[round]{natbib}
\usepackage{xr-hyper}
\usepackage{setspace}
\usepackage[hidelinks,hypertexnames=false,colorlinks=true,allcolors=blue]{hyperref}

\makeatletter
\def\singlespace{\def\baselinestretch{1}\@normalsize}

\@addtoreset{equation}{section}

\def\singlespace{\def\baselinestretch{1}\@normalsize}

\makeatother

\def\P{\mathbb{P}}
\def\E{\mathbb{E}}
\def\Var{\mbox{Var}}
\def\Cov{\mbox{Cov}}
\def\o{o}
\def\O{\mathcal{O}}

\def\vec{\text{vec}}
\def\c{\text{c}}
\def\m{\text{m}}

\newtheorem{theorem}{Theorem}[section]
\newtheorem{corollary}{Corollary}[section]

\numberwithin{equation}{section}
\begin{document}

	\title{\bf Inference Theory for Functional Covariance Structure Across Sampling Regimes}
	\author{Leheng Cai$^1$$^*$, Weijia Li$^1$$^*$ and Qirui Hu$^2$$^\dagger$
		\\~\\
		{\small \it $^{1}$  Department of Statistics and Data Science, Tsinghua University}
		% \\{\small \it Tsinghua University, Beijing, China}
		\\{\small \it $^{2}$ School of Statistics and Data Science,
Shanghai University of Finance and Economics}\\
{\small \it Department of Mathematics, Ruhr University Bochum}
\\{\small \it $^{*}$ The first two authors contributed equally to this article and are listed in alphabetical order.}
\\{\small \it $^{\dagger}$ Corresponding author: huqirui@mail.shufe.edu.cn}
		% \\{\small \it Shanghai University of Finance and Economics, Shanghai 200433, China}
	}
	\date{}

	\maketitle

	\textbf{Abstract:}
Simultaneous inference for functional covariance and eigensystems is challenging when trajectories are observed with noise at irregular and varying frequencies. We develop a unified framework based on a tensor-product
B-spline covariance estimator fitted to off-diagonal residual
products. Decomposing the subject-level scores according to
overlap among observation pairs yields Gaussian approximations
for the covariance surface that accommodate heterogeneous
observation counts and characterize the transition from sparse
to dense sampling. Spectral perturbation analysis further gives
asymptotic normality for fixed-index eigenvalues and simultaneous
confidence bands for fixed-index eigenfunctions. We further establish joint Gaussian approximations for a
growing collection of eigensystems, and show that estimating the mean from the same
observations preserves the oracle Gaussian approximations.  For implementation, the spline
representation reduces spectral computation to a finite-dimensional
matrix eigendecomposition. A subject-cluster
sandwich estimator further  supports asymptotic calibration through
first-order spectral derivatives or direct Gaussian perturbation.
Synthetic data experiments  confirm theoretical correctness, and real data applications to longitudinal bilirubin
measurements and near-infrared spectra further illustrate the proposed methods.

	\textbf{Keywords:} covariance function; functional data;
Gaussian approximation;   phase transition; simultaneous confidence region.

\setstretch{1.5}
\section{Introduction}
Covariance operators organize much of functional data analysis: they describe
within-trajectory dependence, determine functional principal components, and
support dimension reduction, prediction, and regression
\citep{RiceSilverman1991,RamsaySilverman2005,hsing2015theoretical,
WangChiouMuller2016}.  For fully observed trajectories, the empirical
covariance is an average of subject-level outer products, and classical
operator perturbation yields the asymptotic behavior of its eigensystem
\citep{Dauxois1982AsymptoticTF,Hall2006OnPO}.  In longitudinal studies and many
modern functional experiments, however, each trajectory is observed with noise
at a finite and often irregular collection of points.

The frequency of within-subject observations plays a fundamental role in determining the asymptotic behavior of functional estimators.  Sparse designs
require information to be pooled across subjects, as in covariance smoothing
and functional principal component analysis (FPCA) \citep{yao2005functional,HallMullerWang2006},
whereas sufficiently dense designs can recover complete-trajectory behavior
\citep{ZhangChen2007}.  General uniform rates for functional estimators confirm
that the number of measurements per subject is an essential asymptotic index
\citep{Li2010UniformCR}.  {These phase transitions depend on the inferential target:
fixed-index eigenvalues may achieve root-$n$ convergence
while eigenfunction estimation still exhibits nonparametric
rates, and the transition thresholds for mean, covariance,
and spectral estimation need not coincide
\citep{HallMullerWang2006,tonycai2011,ZW16,ZhouWeiYao2025,yan2026functional}.}

The sparse-to-dense transition poses particular challenges
for simultaneous inference.  Existing
bands cover mean functions in several sampling regimes
\citep{Degras2011,MaYangCarroll2012,zheng2014smooth}.
\citet{cai2024sparse} established a unified Gaussian approximation for
mean-function bands and characterized their sparse-to-dense phase transitions.
Related theory for functional time series covers structural break detection
and inference on jump magnitudes \citep{CaiHu2024Breaks}, as well as
self-normalized tests of relevant hypotheses across sampling frequencies
\citep{CaiHu2025Relevant}.
Covariance bands have been developed for dense functional data
\citep{cao2016oracle,wang2020simultaneous}. For spectral inference under dense sampling, \citet{CAI2024107900,CH23}
developed individual and global inference for eigensystems of functional
and imaging data, while \citet{caiHuFPCscore} constructed simultaneous bands
and goodness-of-fit tests for principal component score distributions.

{Extending simultaneous inference for covariance functions and
their eigensystems from dense to sparse functional data poses
several challenges. For noisy and irregularly observed
trajectories, covariance estimation requires smoothing a
bivariate surface using off-diagonal products that are
dependent within subjects. Two such products may share
two, one, or no observation indices, and the resulting
variance contributions have different orders as the
within-subject sampling frequency changes. Heterogeneous
observation counts further affect their relative magnitudes,
making it necessary to account for all three contributions
across sparse-to-dense sampling regimes.

Additional difficulties arise in spectral inference.
Although eigenvalue fluctuations are characterized by
linear projections of the covariance perturbation,
eigenfunction fluctuations involve reduced resolvents
and are sensitive to small eigengaps. Simultaneous
inference over multiple components therefore requires
uniform control of spectral perturbation errors while
preserving the dependence among eigenvalues and
eigenfunctions induced by their common covariance
estimator. Estimating the mean from the same observations
introduces further dependence through the residual products.

We address these challenges through the residualized subject-level
scores of a tensor-product B-spline covariance estimator.
Decomposing their covariance according to the three pair-overlap
patterns yields a unified Gaussian approximation for the covariance
surface across sparse-to-dense sampling regimes, accommodating
heterogeneous observation counts. This decomposition also
identifies how the dominant variance contributions change with
sampling frequency. We further establish Gaussian approximations
for fixed-index eigenvalues and eigenfunctions by applying
spectral derivatives to the covariance perturbation. These
results yield asymptotic normality for individual eigenvalues
and simultaneous confidence bands for individual eigenfunctions,
while characterizing their sparse-to-dense phase
transitions.

We then extend the analysis to joint inference over a diverging
number of eigensystem components. By combining uniform spectral perturbation bounds
with spectral variance bounds and applying spectral derivatives
to the common approximating Gaussian process, we establish joint Gaussian
approximations for eigenvalues and eigenfunctions, simultaneously
over component indices and functional arguments.  Finally, we show that estimating the mean from the
same observations preserves the oracle Gaussian approximations
under suitable conditions.
}

The tensor-product B-spline representation reduces the eigenproblem to
a matrix formed from the spline coefficients and basis Gram matrix.
Its dimension is the number of basis functions and does not increase
when the integration grid is refined.
For implementation, we construct a subject-cluster sandwich estimator
and use first-order spectral perturbation expansions according to the developed Gaussian approximation theorems to obtain critical values for eigenvalue confidence intervals
and simultaneous eigenfunction bands. We also propose a direct Gaussian
perturbation approach that recomputes the eigensystem under simulated
covariance perturbations.

The rest of the paper is organized as follows. Section~\ref{sec:main-results} develops the
statistical framework and theory, Section~\ref{sec:implement} describes implementation,
Section~\ref{sec:simulation} presents {simulation studies}, and Section~\ref{sec:application}  illustrates two applications. 
The supplement gives additional simulation results, technical lemmas and proofs of the main results.

\section{Statistical framework and inference}\label{sec:main-results}
We begin with the observation model and first study an oracle covariance
estimator that uses the true mean.  Then, the feasible estimator in
Section~\ref{sec:feasibleCov} fits the mean from the same observations.
Throughout, asymptotics are indexed by the number of independent subjects
$n$; $N_i$, $J_\c$, and $J_\m$ may depend on $n$.

For subject $i$, let $X_{ij}$ be i.i.d. design points with density $f$ and
observe
\begin{align}\label{eq:observation-model}
Y_{ij}
&=m(X_{ij})+\sum_{k=1}^{\infty}
 \lambda_k^{1/2}\xi_{ik}\psi_k(X_{ij})
 +\sigma(X_{ij})\epsilon_{ij},
\quad 1\le i\le n,\quad 1\le j\le N_i.
\end{align}
Here $\{(\lambda_k,\psi_k):k\ge1\}$ denotes the positive eigensystem of
$\mathcal G$, ordered so that $\lambda_1\ge\lambda_2\ge\cdots>0$; the
positive eigenfunctions form an orthonormal basis of
$\overline{\operatorname{ran}(\mathcal G)}$.  We write $P_0$ for the
orthogonal projection onto $\ker(\mathcal G)$.  The standardized scores satisfy
$\E(\xi_{ik})=0$ and
$\E(\xi_{ik}\xi_{i\ell})=\mathbf 1(k=\ell)$.  The score vectors are i.i.d. across subjects, and their coordinates
need not be independent.  The scores, design points, and i.i.d. errors
$\epsilon_{ij}$ are mutually independent, with $\E(\epsilon_{ij})=0$ and
$\Var(\epsilon_{ij})=1$, and $\sigma^2(x)$ is the conditional measurement-error
variance.  Observation counts $\{N_i\}_{i=1}^{n}$ form a deterministic triangular array.  The latent centered
trajectory is
$\xi_i(x)=\sum_{k\ge1}\lambda_k^{1/2}\xi_{ik}\psi_k(x)$, with covariance
$G(x,x')=\sum_{k\ge1}\lambda_k\psi_k(x)\psi_k(x')$.

\paragraph{Notation}
For $\bm a=(a_1,\ldots,a_s)^\top\in\mathbb R^s$, let
$\|\bm a\|_r=(\sum_{j=1}^s|a_j|^r)^{1/r}$ for $1\le r<\infty$ and
$\|\bm a\|_\infty=\max_{1\le j\le s}|a_j|$.
For an $n\times n$ matrix $\mathbf A$, let
$\|\mathbf A\|_r=\sup_{\|\bm x\|_r=1}\|\mathbf A\bm x\|_r$ for
$1\le r\le\infty$, with $\|\mathbf A\|_{op}=\|\mathbf A\|_2$.  The symbol
$\mathbf A\otimes\mathbf B$ denotes a Kronecker product.  For a random
variable $X$, set $\|X\|_r=\{\E|X|^r\}^{1/r}$.  For a function $h$ on
$\mathcal D$, write
$\|h\|_\infty=\sup_{x\in\mathcal D}|h(x)|$ and
$\|h\|_{\mathcal L^2}=\{\int_{\mathcal D}h^2(x)\,dx\}^{1/2}$.
For \(\mathcal D\subset\mathbb R^d\), a nonnegative integer \(q\), and
\(\nu\in(0,1]\), let \(\mathcal H_{q,\nu}(\mathcal D)\) be the isotropic
H\"older class of functions whose partial derivatives through order \(q\) are
bounded and for which
\(\max_{\|\bm\alpha\|_1=q} \sup_{\substack{t,s\in\mathcal D\\t\ne s}} \frac{|\partial^{\bm\alpha}h(t)-\partial^{\bm\alpha}h(s)|} {\|t-s\|_2^\nu}<\infty.\)
For \(q=0\), \(\partial^{\bm 0}h=h\).  This convention covers both the
univariate and multivariate H\"older classes used below.
For $a>0$, $[a]$ denotes its integer part.
For real numbers $a$ and $b$, $a\vee b$ denotes $\max\{a,b\}$.
For two sequences $\{a_n\}$ and $\{b_n\}$, we write $a_n\asymp b_n$, $a_n\lesssim b_n$ and $a_n\ll b_n$ to mean
that there exist $c$, $C>0$ such that $0<c\leq |a_n/b_n|\leq C<\infty$ for all $n$,  $a_n\leq  Cb_n$ for all $n$, and $a_n=\o(b_n)$
respectively.
Throughout the manuscript and supplement, \(\o(\cdot)\) denotes
little-\(o\), whereas \(\O(\cdot)\) denotes big-\(O\).
\subsection{B-spline covariance estimator}\label{sec:infeasibleCov}
Assume for the moment that $m$ is known, and put $U_{ij}=Y_{ij}-m(X_{ij})$.
For $j_1\ne j_2$,
$\E(U_{ij_1}U_{ij_2}\mid X_{ij_1},X_{ij_2})=G(X_{ij_1},X_{ij_2})$;
excluding the diagonal therefore removes the additive measurement-error
variance without requiring a preliminary estimate of $\sigma^2(\cdot)$.  We regress
these off-diagonal products on a tensor-product spline basis.  
% Ordered pairs
% are used in the formulas below; this convention accounts for the factors
% $1/2$ and $1/4$ in the variance decomposition.

To describe the spline functions, let $\{t_l\}_{l=0}^{J+1}$ be a sequence of equally-spaced points, where $t_l = l / (J+1)$ for $0 \leq l \leq J+1$, which divide $[0,1]$ into $J+1$ equal sub-intervals, denoted as $I_l = [t_l, t_{l+1})$ for $l = 0, \ldots, J-1$, and $I_{J} = [t_{J}, 1]$. Let $\mathcal{S}^{p}_{J} = \mathcal{S}^{p}_{J}[0,1]$ be the polynomial spline space of order $p$ over $
\left\{I_l\right\}_{l=0}^{J}$, consisting of all functions that are $(p-2)$ times continuously differentiable on $[0,1]$ and are polynomials of degree $(p-1)$ within the sub-intervals $I_l$, $l = 0,\ldots,J$. With a {slight} abuse of notation, let $\{B_{l,p}(x)\}_{1 \leq l \leq J + p}$ be the $p$-th order B-spline basis functions of $\mathcal{S}_{J}^{p}$, then $\mathcal{S}_{J}^{p}=\left\{
 \sum_{l =1}^{J+p}c_lB_{l,p}(x)
: c_l\in \mathbb{R} \right\} $.
The tensor-product B-spline estimator is the function
\begin{align*}
    \widetilde G
    =\mathop{\arg\min}_{g\in
    \mathcal{S}_{J_\c}^{p_\c}\otimes\mathcal{S}_{J_\c}^{p_\c}}
    \sum_{i=1}^n\sum_{j_1\neq j_2}^{N_i}
    \{g(X_{ij_1},X_{ij_2})-U_{ij_1}U_{ij_2}\}^2.
\end{align*}
Let $P_{n,2}(N)=n^{-1}\sum_{i=1}^n N_i(N_i-1)$,
$P_{n,3}(N)=n^{-1}\sum_{i=1}^n N_i(N_i-1)(N_i-2)$, and
$P_{n,4}(N)=n^{-1}\sum_{i=1}^n N_i(N_i-1)(N_i-2)(N_i-3)$.
The factorial moments $P_{n,r}(N)$ record the average number of ordered
$r$-tuples contributed by a subject.  They permit heterogeneous observation
counts and determine the weights of the three variance components.  The quantity $\bar{\bar N}=P_{n,2}^{1/2}(N)$ is the effective
within-subject pair scale; it coincides in order with a common $N_i$ under a
balanced design.
For technical convenience, assume that
$\sup_n\max_{1\leq i\leq n}N_i(N_i-1)/P_{n,2}(N)<\infty$ and
$N_i\geq4$.
We write
$\bm B(x)=\bm B_{J_\c,p_\c}(x)=
\left(B_{1,p_\c}(x),\ldots,B_{J_\c+p_\c,p_\c}(x)\right)^\top$.
Throughout the theoretical development, the spline order $p_\c$ is fixed and
the knot sequence is quasi-uniform: if $0=t_0<t_1<\cdots<t_{J_\c+1}=1$ are
the boundary and interior knots, then, for constants independent of $n$, $cJ_\c^{-1}\le t_{j+1}-t_j\le CJ_\c^{-1}$ for $0\le j\le J_\c$.
The mean spline estimator in Section \ref{sec:feasibleCov} has the fixed-order $p_\m$, quasi-uniform geometry with $J_\m$
in place of $J_\c$. 
The tensor-product B-spline basis is then defined by
$\bm B(x,x^\prime)=\bm B(x)\otimes \bm B(x^\prime)$.
Let $\mathbf Q = \E\bm B(X,X^\prime)\bm B^\top(X,X^\prime)$, where $X$ and $X^\prime$ are i.i.d. copies of $X_{ij}$, and \begin{align*}
    \widehat{\mathbf{Q}}= \frac{1}{\sum_{i=1}^{n}N_i(N_i-1)}\sum_{i=1}^{n}\sum_{j_1\neq j_2}^{N_i} \bm B(X_{ij_1},X_{ij_2}) \bm B^\top (X_{ij_1},X_{ij_2}).
\end{align*}
Then
\begin{align*}
    \widetilde G(x,x^\prime)= \bm B^{\top}(x,x^\prime)\widehat{\mathbf Q}^{-1}\left\{\frac{1}{\sum_{i=1}^{n}N_i(N_i-1)}\sum_{i=1}^{n}\sum_{j_1\neq j_2}^{N_i} \bm B (X_{ij_1},X_{ij_2})U_{ij_1}U_{ij_2}\right\}.
\end{align*}

Let $\Gamma(x_1,x_2,x_3,x_4)=\E\{\xi(x_1)\xi(x_2)\xi(x_3)\xi(x_4)\}$ and
\begin{align*}
   \Omega(x_1,x_2,x_3)&=\E\left\{U_{i1}^2U_{i2}U_{i3}|X_{i1}=x_1,X_{i2}=x_2,X_{i3}=x_3\right\} \\&= \E\left\{\xi^2(x_1)\xi(x_2)\xi(x_3) \right\}+\sigma^2(x_1)G(x_2,x_3).
\end{align*}
  The theoretical results
use the following conditions.
\begin{itemize}
    \item[(A1)] For some integer $q>0$ and $\mu\in(0,1]$,
    $G\in\mathcal H_{q,\mu}([0,1]^2)$.  Write $q^\ast=q+\mu$.
\item[(A2)]  For some $\beta_1\in(0,1]$ and any $x\in[0,1]$,   $\Omega(x,\cdot,\cdot)\in\mathcal{H}_{0,\beta_1}[0,1]^2$, $\sup_{x\in[0,1]}\left\|\Omega(x,\cdot,\cdot)\right\|_{0,\beta_1}<\infty$, and there exist positive constants $c_{\Omega}$ and $C_{\Omega}$ such that  $c_{\Omega}\le\inf_{x,x^\prime}\Omega(x,x^\prime,x^\prime)\le\sup_{x,x^\prime}\Omega(x,x^\prime,x^\prime)\le C_{\Omega}$.
\item[(A3)] For some $\beta_2\in(0,1]$,    $\Gamma(\cdot,\cdot,\cdot,\cdot)\in\mathcal{H}_{0,\beta_2}[0,1]^4$, and    there exist positive constants $c_{\Gamma}$ and $C_{\Gamma}$ such that  $c_{\Gamma}<\inf_{x,x^\prime}\{\Gamma(x,x,x^\prime,x^\prime)-G^2(x,x')\}<\sup_{x,x^\prime}\Gamma(x,x,x^\prime,x^\prime)<C_{\Gamma}$.
\item[(A4)] For some positive constants $c_U$ and $C_U$,  $c_U\le\inf_{x_1,x_2\in[0,1]}\E(U_{i1}^2U_{i2}^2|X_{i1}=x_1,X_{i2}=x_2)\le\sup_{x_1,x_2\in[0,1]}\E(U_{i1}^2U_{i2}^2|X_{i1}=x_1,X_{i2}=x_2)\le C_U$

		\item[(A5)] For some positive constants $c_{f}$ and $C_{f}$, $c_{f}\leq \inf_{x\in[0,1]} f(x)  \leq \sup_{x\in[0,1]} f(x) \leq C_{f}$
  \item[(A6)]
  The measurement-error scale $\sigma(\cdot)$ is uniformly bounded, and for
some integers $r_1,r_2>2$,
$\E|\epsilon_{ij}|^{2r_1}<\infty$ and
$\sup_{x\in[0,1]}\E|\xi_i(x)|^{2r_2}<\infty$.
\item[(A7)] Let  $r=\min\{r_1,r_2\}$, the spline order $p_\c \geq q^\ast$, and
$$\left(\bar{\bar{N}}^{-2}J_\c^2+ \bar{\bar{N}}^{-1}J_\c +1\right)J_\c^{-4}\left\{n/\log(n)\right\}^{1-2/r}\to\infty.$$
\begin{itemize}
    \item[(1)] When $\bar{\bar{N}}\lesssim J_\c $, $J_\c^{-q^\ast-1}n^{1/2}\bar{\bar{N}} \log^{1/2}(n)=\o(1)$,
     $  J_\c^{2(3-4/r)}n^{-1+2/r} \log^2(n)\bar{\bar{N}}^{ -2+4/r }
      =\o(1).$
  \item[(2)]When $\bar{\bar{N}}\gg J_\c$, $J_\c^{-q^\ast}n^{1/2}\log^{1/2}(n)=\o(1)$,
      $  J_\c^{2(2-2/r)}n^{-1+2/r}\log^2(n)=\o(1).$
\end{itemize}

\end{itemize}

Assumption (A1) controls approximation of the covariance surface.  
Assumptions (A2)--(A4) are tied to the three overlap patterns.  The
three-argument function \(\Omega\) enters when two pair products share one
measurement, whereas the four-argument function \(\Gamma\) governs disjoint
pairs from the same trajectory.  Their H\"older conditions permit uniform
spline approximation of the corresponding covariance kernels, and the lower
bounds prevent the pointwise standardization from degenerating.  Assumption
(A4) plays the analogous role for a single observed residual product and
therefore contains both latent-trajectory and measurement-error variation.
Assumption (A5) keeps the design density positive and the tensor Gram matrix
well conditioned at its natural \(J_\c^{-2}\) scale.  Assumption (A6) bounds
the moments needed for truncation and Gaussian approximation over a growing spline basis. The restriction of parameters is given in (A7).

For use in the residualized score, define
\(\Omega_c(x_1,x_2,x_3) =\Omega(x_1,x_2,x_3)-G(x_1,x_2)G(x_1,x_3),\) \(\Gamma_c(x_1,x_2,x_3,x_4) =\Gamma(x_1,x_2,x_3,x_4)-G(x_1,x_2)G(x_3,x_4).\)
Assumptions (A1)--(A4) provide the required regularity and upper bounds.  In
particular, (A3) gives a uniform positive lower bound for
$\Gamma_c(x,x',x,x')$ and hence for the two-index residual variance, while
$\Omega_c(x,x',x')\ge \Gamma_c(x,x',x,x')$.
The population least-squares coefficient is
\(
\bm\theta=\mathbf Q^{-1}\E\{\bm B(X,X')G(X,X')\}.
\)
Let $\widetilde{\bm B}(x,x')=\bm B(x,x')+\bm B(x',x)$
 and \(\widetilde{\bm Z}^{\circ}_{i,ab} =\widetilde{\bm B}(X_{ia},X_{ib}) \{U_{ia}U_{ib} -\bm B^\top(X_{ia},X_{ib})\bm\theta\}.\)
Writing $\bm B_{i,ab}=\bm B(X_{ia},X_{ib})$, the centered product moment
satisfies
\begin{equation*}
\bm B_{i,ab}U_{ia}U_{ib}
-\E\{\bm B(X,X')G(X,X')\}
=
\bm B_{i,ab}\{U_{ia}U_{ib}-\bm B_{i,ab}^\top\bm\theta\}
+\{\bm B_{i,ab}\bm B_{i,ab}^{\top}-\mathbf Q\}\bm\theta.
\end{equation*}
Further define
\(
R_{i,j_1j_2}
=U_{ij_1}U_{ij_2}
-\bm B^\top(X_{ij_1},X_{ij_2})\bm\theta.
\)
The sample normal equations then give
\begin{equation*}
\widetilde G(x,x')-\bm B^\top(x,x')\bm\theta
=\bm B^\top(x,x')\widehat{\mathbf Q}^{-1}
 \frac{1}{nP_{n,2}(N)}
 \sum_{i=1}^n\sum_{j_1\ne j_2}^{N_i}
 \bm B(X_{ij_1},X_{ij_2})R_{i,j_1j_2}.
\end{equation*}
Consequently, the covariance of the linear term is determined by the residual
score.  Grouping ordered pairs into
symmetrized unordered-pair scores, write
$\mathbf C_{ab,cd}=\Cov(\widetilde{\bm Z}^{\circ}_{i,ab},
\widetilde{\bm Z}^{\circ}_{i,cd})$.  Then $\mathbf\Sigma=\mathbf\Sigma_1+\mathbf\Sigma_2+\mathbf\Sigma_3$, where
\[
\mathbf\Sigma_1=\frac{1}{2P_{n,2}(N)}\mathbf Q^{-1}\mathbf C_{12,12}\mathbf Q^{-1},\;
\mathbf\Sigma_2=\frac{P_{n,3}(N)}{P_{n,2}^2(N)}\mathbf Q^{-1}\mathbf C_{12,13}\mathbf Q^{-1},\;
\mathbf\Sigma_3=\frac{P_{n,4}(N)}{4P_{n,2}^2(N)}\mathbf Q^{-1}\mathbf C_{12,34}\mathbf Q^{-1}.
\]
The three matrices correspond to distinct sources of variation.  The
two-index component $\mathbf\Sigma_1$ is driven by a single residual product
and has pointwise order $J_\c^2/P_{n,2}(N)$.  The shared-index component
$\mathbf\Sigma_2$ retains dependence created by one common observation and has
order $J_\c P_{n,3}(N)/P_{n,2}^2(N)$.  The disjoint-pair component
$\mathbf\Sigma_3$ reflects the common latent trajectory within a subject and
has order $P_{n,4}(N)/P_{n,2}^2(N)$.  Which term dominates is determined by
the relation between $\bar{\bar N}$ and the spline resolution $J_\c$. 
% The
% cluster sandwich in \eqref{eq:cluster-sandwich} computes their sum without an
% explicit quadruple loop, because all off-diagonal pair scores from one subject
% are aggregated before the outer product is formed.
Define a Gaussian process $\mathcal{Z}_n^\ast(x,x^\prime)=\mathcal{Z}_n(x,x^\prime)/\{\Var(\mathcal{Z}_n(x,x^\prime))\}^{1/2}$, in which $\mathcal{Z}_n(x,x^\prime)$ has mean zero and covariance function ${\text {Cov}}\left(\mathcal{Z}_n (x,x^\prime),\mathcal{Z}_n (y,y^\prime)\right)=\bm B^\top(x,x^\prime){\boldsymbol\Sigma}\bm B(y,y^\prime)$.

The first theorem gives a unified Gaussian approximation. 
\begin{theorem}\label{theorem1}
Under Assumptions (A1)--(A7), the following simultaneous approximation holds.
More precisely, on an extension of the probability space one can construct
\(\bm G_n\sim N\{\bm0,\mathbf I_{(J_\c+p_\c)^2}\}\) such that
\[
\sup_{x,x'\in[0,1]}
\left|
\frac{\sqrt n\{\widetilde G(x,x')-G(x,x')\}}
     {\|\boldsymbol\Sigma^{1/2}\bm B(x,x')\|_2}
-
\frac{\bm B^\top(x,x')\boldsymbol\Sigma^{1/2}\bm G_n}
     {\|\boldsymbol\Sigma^{1/2}\bm B(x,x')\|_2}
\right|
=\o_p\{(\log n)^{-1/2}\}.
\]
The coupling holds in \(\mathcal C([0,1]^2)\). Then, 
  \begin{align}\label{semi-dense}
    \sup_{t\in\mathbb{R}}\left|\P\left(\sup_{x,x^\prime\in[0,1]}\frac{\sqrt{n}\left|\widetilde G(x,x^\prime)-G(x,x^\prime)\right|}{\left\|{\boldsymbol \Sigma}^{1/2}\bm B(x,x^\prime)\right\|_2}\leq t\right)- \P\left(\sup_{x,x^\prime\in[0,1]}\left|\mathcal{Z}_n^\ast(x,x^\prime)\right| \leq t\right)\right|\to 0.
\end{align}
\end{theorem}

Further define   Gaussian processes  $\mathcal{Z}_{1n}^\ast(x,x^\prime) = \mathcal{Z}_{1n}(x,x^\prime)/\{\Var(\mathcal{Z}_{1n}(x,x^\prime))\}^{1/2}$ (depending on $n$) and $\mathcal{Z}_3^\ast(x,x^\prime) = \mathcal{Z}_{3}(x,x^\prime)/\{\Var(\mathcal{Z}_{3}(x,x^\prime))\}^{1/2}$, where $\mathcal{Z}_{1n}(x,x^\prime)$ and $\mathcal{Z}_{3}(x,x^\prime)$ have mean zero and covariance structure as follows, \begin{align*}
    &{\text {Cov}}\left(\mathcal{Z}_{1n} (x,x^\prime),\mathcal{Z}_{1n} (y,y^\prime)  \right)=  \bm B^\top(x,x^\prime) {\boldsymbol \Sigma}_1\bm B(y,y^\prime),\\
    &{\text {Cov}}\left(\mathcal{Z}_{3}  (x,x^\prime),\mathcal{Z}_{3}  (y,y^\prime)  \right)=\Gamma(x,y,x^\prime,y^\prime)-G(x,x^\prime)G(y,y^\prime).
\end{align*}
\begin{theorem}\label{theorem2}
Suppose that Assumptions (A1)-(A7) hold. \begin{itemize}
\item[(1)] When $\bar{\bar{N}}\log^2(n)\ll J_\c$, as $n\to\infty$, 
\begin{align}\label{sparse}
    \sup_{t\in\mathbb{R}}\left|
    \P\left(\sup_{x,x'\in[0,1]}\left|\frac{\sqrt n\{\widetilde G(x,x')-G(x,x')\}}
{\|\boldsymbol\Sigma_1^{1/2}\bm B(x,x')\|_2}\right|\leq t\right)
    -\P\left(\sup_{x,x'\in[0,1]}|\mathcal{Z}_{1n}^\ast(x,x')|\leq t\right)
    \right|\to 0.
\end{align}
\item[(2)]
When $\bar{\bar N}\log^{-1}(n)\gg J_\c$ and 
$ {P_{n,4}(N)}/{P_{n,2}^2(N)}\to1$
as \(n\to\infty\), 
\begin{align}\label{dense}
    \sup_{t\in\mathbb{R}}\left|
    \P\left(\sup_{x,x'\in[0,1]}\left| \frac{\sqrt n\{\widetilde G(x,x')-G(x,x')\}}
{\{\Gamma(x,x,x',x')-G^2(x,x')\}^{1/2}}\right|\leq t\right)
    -\P\left(\sup_{x,x'\in[0,1]}|\mathcal{Z}_{3}^\ast(x,x')|\leq t\right)
    \right|\to 0.
\end{align}
\end{itemize}

\end{theorem}

In the sparse region, pair-level variation dominates and the limiting
correlation structure continues to depend on the spline resolution.  In the
dense region, each subject has enough measurements for the disjoint-pair
term to recover the fluctuation of the latent subject-level covariance.  The
condition $P_{n,4}(N)/P_{n,2}^2(N)\to1$ is automatic for a common increasing
count; without it, the
dense covariance is multiplied by the nonunit limiting count factor. 
% Between
% these extremes the full $\mathbf\Sigma_1+\mathbf\Sigma_2+\mathbf\Sigma_3$
% approximation in Theorem~\ref{theorem1} remains necessary.
{Corollary~\ref{COR:phasetransition} gives phase-transition  regimes from sparse to dense under corresponding knot choices.} 
\begin{corollary}\label{COR:phasetransition}

Suppose that Assumptions (A1)-(A7) hold. For any fixed $b>1/(2q^\ast)$,
\begin{itemize}
    \item[(1)]
    When $\bar{\bar{N}}\ll n^{1/(2q^\ast)}\log^{b-2(q^\ast+1)/q^\ast}(n)$ and  $J_\c\asymp \bar{\bar{N}}^{1/(q^\ast+1)}n^{1/(2q^\ast+2)}\log^{bq^\ast/(q^\ast+1)}(n)$,
    (\ref{sparse}) holds; this regime is called ``sparse''.
     \item[(2)]
     When $n^{1/(2q^\ast)}\log^{b-2(q^\ast+1)/q^\ast}(n)\lesssim \bar{\bar{N}}\lesssim n^{1/(2q^\ast)}\log^{b+1}(n)$ and  $J_\c\asymp  n^{1/(2q^\ast )}\log^{b}(n)$,
    (\ref{semi-dense}) holds; this regime is called ``semi-dense''.
\item[(3)]
When $\bar{\bar{N}}\gg n^{1/(2q^\ast)}\log^{b+1}(n)$,
$n^{1/(2q^\ast)}\ll J_\c\ll\bar{\bar N}\log^{-1}(n)$, and
$P_{n,4}(N)/P_{n,2}^2(N)\to1$,
    (\ref{dense}) holds; this regime is called ``dense''.
\end{itemize}
\end{corollary}
We note that in assertion (2) of Theorem \ref{theorem2} and assertion (3) of Corollary \ref{COR:phasetransition}, the condition $J_\c^{-q^\ast}n^{1/2}\log^{1/2}(n)=\o(1)$ in Assumption (A7) can be modified   to $J_\c^{-q^\ast}n^{1/2}=\o(1)$.

\subsection{Eigenvalues}\label{sec:FPCA}
Let
$\widetilde G_s(x,u)=\{\widetilde G(x,u)+\widetilde G(u,x)\}/2.$
Let $\mathcal G$ and $\widetilde{\mathcal G}$ be the integral operators with
kernels $G$ and $\widetilde G_s$, respectively:
$(\mathcal Gh)(x)=\int_0^1G(x,u)h(u)\,du$ and
$(\widetilde{\mathcal G}h)(x)=\int_0^1\widetilde G_s(x,u)h(u)\,du$.
Denote their fixed-index eigenpairs by $(\lambda_k,\psi_k)$ and
$(\widetilde\lambda_k,\widetilde\psi_k)$.  The fitted operator is
self-adjoint and finite rank but need not be positive semidefinite; the theory
below concerns an isolated positive population eigenvalue and its empirical
continuation.
Assume that the non-zero eigenvalues are strictly decreasing:
$\lambda_1>\lambda_2>\lambda_3>\cdots$.  For each empirical eigenfunction,
choose its sign so that
$\langle\widetilde\psi_k,\psi_k\rangle_{\mathcal L^2}\geq0$.

For the fixed positive eigenvalue $\lambda_k$, let
$\bm b_{kk}=\iint_{[0,1]^2}\bm B(u,v)\psi_k(u)\psi_k(v)\,du\,dv$.
The exact variance of the finite-spline least-squares linear influence
functional is
\(\sigma_{nk}^2=\bm b_{kk}^\top\mathbf\Sigma\bm b_{kk}\).
{Under the conditions of Theorem~\ref{THM:eigenvalues}}, it has
the componentwise asymptotic expansion
\begin{equation}\label{eq:eigenvalue-three-component-variance}
\begin{aligned}
\sigma_{nk}^2
={}&\frac{2}{P_{n,2}(N)}\Bigg[
 \E 
 \left\{\int_0^1\frac{\psi_k^2(x)}{f(x)}
 \{\xi_i^2(x)+\sigma^2(x)\}\,dx\right\}^{\!2}
 -\iint_{[0,1]^2}
 \frac{\psi_k^2(x)\psi_k^2(y)}{f(x)f(y)}
 G^2(x,y)\,dx\,dy\Bigg]\\
&+\frac{4P_{n,3}(N)}{P_{n,2}^2(N)}\left(
 \lambda_k\E\!\left[
 \xi_{ik}^2\int_0^1\frac{\psi_k^2(x)}{f(x)}
 \{\xi_i^2(x)+\sigma^2(x)\}\,dx
 \right] 
 -\lambda_k^2\int_0^1\frac{\psi_k^4(x)}{f(x)}\,dx\right)\\
&+\frac{P_{n,4}(N)}{P_{n,2}^2(N)}
 \lambda_k^2\{\E(\xi_{ik}^4)-1\}
 +\mathfrak r_{nk},
\end{aligned}
\end{equation}
where, with {$\beta_3$ and $\beta_4$ as defined in (B1) below and}
\(
\beta_\star=\min\{\beta_3,\beta_4\}, 
\delta_{J_\c}=J_\c^{-\beta_\star}+J_\c^{-q^\ast},
\)
there is a constant \(C_k\), independent of \(n\), such that
\[
|\mathfrak r_{nk}|
\leq C_k\delta_{J_\c}\left\{
P_{n,2}^{-1}(N)
+P_{n,3}(N)P_{n,2}^{-2}(N)
+P_{n,4}(N)P_{n,2}^{-2}(N)
\right\}.
\]
Thus,
\(\mathfrak r_{nk}=o\!\left\{ P_{n,2}^{-1}(N) +P_{n,3}(N)P_{n,2}^{-2}(N) +P_{n,4}(N)P_{n,2}^{-2}(N) \right\}.\)
Each of the three leading overlap contributions in
\eqref{eq:eigenvalue-three-component-variance} is shown to be nonnegative.  The negative terms displayed in the first two brackets are
the centering corrections generated by the random-design least-squares
normal equation. 

{
\citet[Theorem~2]{ZhouWeiYao2025} establish the asymptotic
normality of the eigenvalue  $\widehat\lambda_k^{\mathrm{ker}}$ for a local-linear covariance estimator with subject
weights $\{nN_i(N_i-1)\}^{-1}$. Under the dense-sampling and
negligible-bias conditions of their Corollary~4(a), their 
result (expressed in our  notation) is
\[
\sqrt{n}(\widehat\lambda_k^{\mathrm{ker}}-\lambda_k)
\xrightarrow{d}
\mathcal N\!\left(
0,24\lambda_k^2\{\E(\xi_{ik}^4)-1\}
\right).
\]
 In contrast, under the dense conditions of
Corollary~\ref{COR:eigenvalues}, the asymptotic variance of $\widehat\lambda_k$ in
\eqref{eq:eigenvalue-three-component-variance} yields
$\lambda_k^2\{\E(\xi_{ik}^4)-1\}$.
For Gaussian FPC scores, the two asymptotic variances are
$48\lambda_k^2$ and $2\lambda_k^2$, respectively, which shows that our spline estimator has a theoretical efficiency advantage
in the dense regime.
}

\begin{itemize}
\item[(B1)] For some $\beta_3,\beta_4\in(0,1]$,   $f(\cdot) \in\mathcal H_{0,\beta_3}[0,1]$ and $\psi_k(\cdot)\in\mathcal H_{0,\beta_4}[0,1]$.
    \item[(B2)]   Let  $r=\min\{r_1,r_2\}$, the spline order $p_\c \geq q^\ast$,  $J_\c^{-q^\ast}n^{1/2} =\o(1)$, and
$$\left(\bar{\bar{N}}^{-2}J_\c^2+ \bar{\bar{N}}^{-1}J_\c +1\right)J_\c^{-4}\left\{n/\log(n)\right\}^{1-2/r}\to\infty.$$
\begin{itemize}
\item[(1)] When $\bar{\bar{N}}\lesssim J_\c $,
     $  J_\c^{2(3-4/r)}n^{-1+2/r} \log (n)\bar{\bar{N}}^{ -2+4/r }
      =\o(1).$
  \item[(2)]When $\bar{\bar{N}}\gg J_\c$,
      $  J_\c^{2(2-2/r)}n^{-1+2/r}\log (n)=\o(1).$

\end{itemize}
\item[(B3)] For the fixed index $k$, $\liminf_{n\to\infty}\sigma_{nk}^2>0$.
\end{itemize}

Assumption (B1) is used to approximate the eigenvalue influence functional by
its spline reproducing-kernel counterpart.  Assumption (B2) {imposes 
tuning restrictions} to make the  spectral
perturbation remainder asymptotically negligible.  Assumption
(B3) keeps the fixed-index linearization variance positive. The eigenvalue
influence is a projection of the same residualized least-squares score
used for covariance inference.

\begin{theorem}
    \label{THM:eigenvalues}
    Under Assumptions (A1)--(A6) and (B1)--(B3), as $n\to\infty$,
     \begin{align*}
    \frac{\sqrt{n}\left(\widetilde\lambda_k-\lambda_k\right)}{\sigma_{nk}}\xrightarrow{d}\mathcal N(0,1).
\end{align*}
\end{theorem}
If \(\bar{\bar N}\to\infty\) and
\(P_{n,4}(N)/P_{n,2}^2(N)\to1\), then
\(P_{n,2}^{-1}(N)\to0\) and
\(P_{n,3}(N)P_{n,2}^{-2}(N)\to0\).  Consequently,
\(
\sigma_{nk}^2\to
\lambda_k^2\Var(\xi_{ik}^2)
=\lambda_k^2\{\E(\xi_{ik}^4)-1\}.
\)
Thus, in the following Corollary \ref{COR:eigenvalues}, the dense root-$n$ limit is nondegenerate when
$\Var(\xi_{ik}^2)>0$; for standard Gaussian scores,
the limit variance is $2\lambda_k^2$.

\begin{corollary}\label{COR:eigenvalues}
    Under Assumptions (A1)-(A6) and (B1)-(B2),
    $\bar{\bar N}\to\infty$,
    $\frac{P_{n,4}(N)}{P_{n,2}^2(N)}\to1$ as $n\to\infty$, and
    $\Var(\xi_{ik}^2)>0$,
    \begin{align*}
     {\sqrt{n}\left(\widetilde\lambda_k-\lambda_k\right)} \xrightarrow{d}\mathcal N\left(0,\left(\E\xi_{ik}^4-1\right)\lambda_k^2\right).
\end{align*}
\end{corollary}
% Under these conditions,
% $\sigma_{nk}^2\to\lambda_k^2\Var(\xi_{ik}^2)>0$ and the denominator in
% (B3) tends to one.  Thus (B3) follows, and Theorem~\ref{THM:eigenvalues}
% together with Slutsky's theorem yields the displayed limit.
% The dense expression is the complete-trajectory influence variance for the
% $k$th eigenvalue.  Away from that regime, the full three-component expression
% in \eqref{eq:eigenvalue-three-component-variance} governs the variance.

\subsection{Eigenfunctions}\label{sec:eigenfunctions}
 \begin{itemize}
     \item[(C1)] For some $\varpi\in(0,1]$, $\sum_{\ell=1}^{\infty}\sqrt{\lambda_\ell}\left\|\psi_\ell\right\|_{0,\varpi}<\infty$,  $\sup_{\ell\geq 1}\|\psi_\ell\|_\infty<\infty$, and $$\inf_{u\in[0,1]}
     \left[
       \lambda_k^{-1}
       \E\{\{\xi_i^2(u)+\sigma^2(u)\}\xi_{ik}^2\}
       -\psi_k^2(u)
     \right]>0.$$   When $\bar{\bar N}\gg J_\c$, $$\inf_{x\in[0,1]}\sum_{\ell\neq k}\sum_{\ell'\neq k}\frac{\lambda_k \sqrt{\lambda_\ell\lambda_{\ell'}}\E(\xi_{ik}^2\xi_{i\ell}\xi_{i\ell'})}{(\lambda_k-\lambda_\ell)(\lambda_k-\lambda_{\ell'})}\psi_\ell(x)\psi_{\ell'}(x)
    >0.$$

     \item[(C2)]  Let  $r=\min\{r_1,r_2\}$, the spline order $p_\c \geq q^\ast$, and
$$\left(\bar{\bar{N}}^{-2}J_\c^2+ \bar{\bar{N}}^{-1}J_\c +1\right)J_\c^{-4}\left\{n/\log(n)\right\}^{1-2/r}\to\infty.$$
\begin{itemize}
    \item[(1)] When $\bar{\bar{N}}\lesssim J_\c $, $J_\c^{-q^\ast-1/2}n^{1/2}\bar{\bar{N}}^{1/2} \log^{1/2}(n)=\o(1)$,
     $  J_\c^{7-8/r }n^{-1+2/r} \log^2(n)\bar{\bar{N}}^{ -3+4/r }
      =\o(1).$
  \item[(2)]When $\bar{\bar{N}}\gg J_\c$, $J_\c^{-q^\ast}n^{1/2}\log^{1/2}(n)=\o(1)$,
      $  J_\c^{2(2-2/r)}n^{-1+2/r}\log^2(n)=\o(1).$
\end{itemize}
 \end{itemize}

 Assumption (C1) controls the supremum norm of eigenfunctions and
ensures a nonvanishing dense-limit variance.  Assumption (C2) is stronger than
its eigenvalue analogue because uniform eigenfunction inference must control
the spectral perturbation remainder as a function of $x$, not merely after an
$L^2$ inner product.  

Throughout,
all spectral sums over $\ell\ne k$ are restricted
to the positive eigenvalues of $\mathcal G$.
Define the reduced resolvent by
\[
\mathcal R_k h
=\sum_{\substack{\ell:\lambda_\ell>0\\\ell\ne k}}
 \frac{\langle h,\psi_\ell\rangle}
 {\lambda_k-\lambda_\ell}\psi_\ell
+\lambda_k^{-1}P_0h.
\]
The term $\lambda_k^{-1}P_0h$ accounts for
contributions from the null space. 
The derivative of the eigenfunction map with
respect to the covariance operator is then
characterized by this reduced resolvent.

For
\(\bm b_{\ell k}=\iint_{[0,1]^2}\bm B(u,v)\psi_\ell(u)\psi_k(v)\,du\,dv,\) \[
\bm d_{k,J_\c}(x)=\sum_{\substack{\ell:\lambda_\ell>0\\\ell\ne k}} \frac{\psi_\ell(x)}{\lambda_k-\lambda_\ell}\bm b_{\ell k} +\lambda_k^{-1}\bm b_{0k}(x),
\]
where
\(
\bm b_{0k}(x)
=
\left[
P_0\left\{
\int_0^1 \bm B(\cdot,v)\psi_k(v)\,dv
\right\}
\right](x),
\)
with $P_0$ acting componentwise on the
first argument of $\bm B(\cdot,\star)$.  Let
$\Xi_{k,n}(\cdot)$ be the centered Gaussian process with covariance
\[
\Cov\{\Xi_{k,n}(x),\Xi_{k,n}(x')\}
=\bm d_{k,J_\c}^\top(x)\mathbf\Sigma\bm d_{k,J_\c}(x').
\]

 \begin{theorem}\label{thm:fpcs}
Under Assumptions (A1)--(A6), (B1), and (C1)--(C2),
for any fixed positive-eigenvalue index $k$, there exist coupled versions of the
empirical and Gaussian processes for which
\[
\sup_{x\in[0,1]}
\left|
\frac{\sqrt n\{\widetilde\psi_k(x)-\psi_k(x)\}}
     {\sqrt{\Var\{\Xi_{k,n}(x)\}}}
-
\frac{\Xi_{k,n}(x)}
     {\sqrt{\Var\{\Xi_{k,n}(x)\}}}
\right|
=\o_p\{(\log n)^{-1/2}\}
\quad\text{in }\mathcal C([0,1]).
\]
Consequently,
  \begin{align*}
  \sup_{t\in\mathbb{R}}\left|\P\left(\sup_{x \in[0,1]}\frac{\sqrt{n}\left|\widetilde{\psi}_k (x)- \psi_k(x) \right|}{\sqrt{\Var(\Xi_{k,n}(x))}}\leq t\right)- \P\left( \sup_{x\in[0,1]}\left|\frac{\Xi_{k,n}(x)}{\sqrt{\Var(\Xi_{k,n}(x))}}\right| \leq t\right)\right|\to 0.
\end{align*}
\end{theorem}
The theorem gives a band for the whole function at a fixed component index. 
For a fixed \(k\), let
\(\{\mathcal N_{\ell k}:\ell\ne k\}\) be a centered Gaussian sequence
satisfying $ \Cov(\mathcal N_{\ell k},\mathcal N_{\ell'k})
    =
    \E(\xi_{ik}^2\xi_{i\ell}\xi_{i\ell'})$ for $\ell,\ell'\ne k$.
Define the mean-zero Gaussian process
\[
    \Xi_{k,3}(x)
    =
    \sum_{\ell\ne k}
    \frac{
        \sqrt{\lambda_k\lambda_\ell}
    }{
        \lambda_k-\lambda_\ell
    }
    \mathcal N_{\ell k}\psi_\ell(x),
    \ x\in[0,1].
\]
Let \(\Xi_{k,12,n}(\cdot)\) be a mean-zero
Gaussian process with covariance function
\[
    \Cov\{\Xi_{k,12,n}(x),\Xi_{k,12,n}(x')\}
    =\bm d_{k,J_\c}^{\top}(x)(\bm\Sigma_1+\bm\Sigma_2)
      \bm d_{k,J_\c}(x').
\]

\begin{theorem}
\label{thm:eigenfunction-phase-transition}
Suppose that   Assumptions (A1)-(A6), (B1), and (C1)-(C2)   hold.

\begin{itemize}
\item[(1)] If
$\bar{\bar N}\log(n)\ll J_\c$,
then
\[
\begin{aligned}
    \sup_{t\in\mathbb R}
    \left|
    \P\left(
        \sup_{x\in[0,1]}
        \frac{
            \sqrt n
            |\widetilde\psi_k(x)-\psi_k(x)|
        }{
            \sqrt{\Var\{\Xi_{k,12,n}(x)\}}
        }
        \leq t
    \right)-
    \P\left(
        \sup_{x\in[0,1]}
        \frac{
            |\Xi_{k,12,n}(x)|
        }{
            \sqrt{\Var\{\Xi_{k,12,n}(x)\}}
        }
        \leq t
    \right)
    \right|
    \to 0.
\end{aligned}
\]

\item[(2)]
If $ \bar{\bar N}\log^{-1}(n)\gg J_\c$ and  ${P_{n,4}(N)}/{P_{n,2}^2(N)}
    \to1$,
then
\[
\begin{aligned}
    \sup_{t\in\mathbb R}
    \left|
    \P\left(
        \sup_{x\in[0,1]}
        \frac{
            \sqrt n
            |\widetilde\psi_k(x)-\psi_k(x)|
        }{
            \sqrt{\Var\{\Xi_{k,3}(x)\}}
        }
        \leq t
    \right)
    -
    \P\left(
        \sup_{x\in[0,1]}
        \frac{
            |\Xi_{k,3}(x)|
        }{
            \sqrt{\Var\{\Xi_{k,3}(x)\}}
        }
        \leq t
    \right)
    \right|
    \to0.
\end{aligned}
\]
\end{itemize}
\end{theorem}

The eigenfunction phase transition is inherited from the covariance influence
process after application of the reduced resolvent.  In sparse designs the
two- and shared-index terms remain in the normalization; in dense designs the
limit depends only on subject-level products of principal-component scores.
The dense count-factor condition ${P_{n,4}(N)}/{P_{n,2}^2(N)}
    \to1$ has the same role as in the covariance and
eigenvalue results.

\subsection{Joint inference over eigen-indices}\label{sec:joint-eigensystem}
For dense trajectories, \citet{CAI2024107900} gave simultaneous inference for
a growing collection of eigensystems. {In this subsection, we also consider joint inference over $1\le k\le K_n$, allowing $K_n\to\infty$.}

For a self-adjoint perturbation $\mathcal H$, use the reduced resolvent
from Section~\ref{sec:eigenfunctions} to define
$\mathcal L_k(\mathcal H)=\langle\mathcal H\psi_k,\psi_k\rangle$ and
$\mathcal D_k(\mathcal H)(x)=(\mathcal R_k\mathcal H\psi_k)(x)$.
Let $\delta_{K_n}$ be the minimum full eigengap among the first $K_n$
eigenvalues. All operators act on $\mathcal L^2([0,1],dx)$, and all
inner products are with respect to Lebesgue measure.  

\begin{itemize}
\item[(S1)] The design density satisfies (A5). For a fixed integer
$p_\c\ge4$, $\E\|\xi_i\|_{C^{p_\c}}^8\le C$. The standardized
KL scores are independent over their component index, with
$\sup_{k\ge1}\E|\xi_{ik}|^4\le C$ and
$\inf_{k\ge1}\Var(\xi_{ik}^2)\ge c>0$.
The errors satisfy $\E|\epsilon_{ij}|^8\le C$, and
$\sigma^2(x)\le C$.

\item[(S2)] 
For $i=1,\ldots,n$, $c\bar N\le N_i\le C\bar N$, where
$\bar N=n^{-1}\sum_iN_i$. Constants do not depend on $n$ or $\bar N$.

\item[(S3)] {The positive eigenvalues of the covariance operator satisfy}
\(ck^{-a}\le\lambda_k\le Ck^{-a},\) \(\lambda_k-\lambda_{k+1}\ge ck^{-a-1},\) \(\sup_k\|\psi_k\|_\infty\le C,\) \(a>1,\)
and
\(
 \inf_{k\ge1,\,x\in[0,1]}
       \{G(x,x)-\lambda_k\psi_k(x)^2\}>0.
\)
Let $a_k\in\mathcal L^2[0,1]$, $k\ge1$, be known deterministic
functions fixed before observing the data, with
$\sup_k\|a_k\|_2\le C$. Orient the population eigenfunctions so that
$\inf_k\langle\psi_k,a_k\rangle\ge c>0$, and choose the empirical
signs so that $\langle\widetilde\psi_k,a_k\rangle\ge0$.

\item[(S4)] Use splines of order $p_\c$, with
\begin{equation}
 \begin{gathered}
 J_\c\asymp n^\beta,~ K_n\to\infty,~ K_n\le Cn^\gamma,~
 \frac1{2p_\c}<\beta<\frac16,~
 0<\gamma<
 \frac{\min\{\beta,p_\c\beta-1/2,\,1/2-3\beta\}}{12(a+1)}.
 \end{gathered}
 \label{cmp:eq:tuning}
\end{equation}
\end{itemize}

The residual variance condition in (S3) requires that
no single principal component accounts for all
pointwise variation, uniformly over the domain.
When $\inf_{x\in[0,1]}G(x,x)>0$, the boundedness of
the eigenfunctions and the decay of the eigenvalues
reduce its verification to finitely many components.

Use the covariance matrix $\mathbf\Sigma=\mathbf\Sigma_1+
\mathbf\Sigma_2+\mathbf\Sigma_3$ from Section~\ref{sec:infeasibleCov}
and the vectors $\bm b_{kk}$ and $\bm d_{k,J_\c}(x)$ from
Sections~\ref{sec:FPCA} and~\ref{sec:eigenfunctions}.
For a vector $\bm G_n\sim N\{\bm0,\mathbf I_{(J_\c+p_\c)^2}\}$,
define the Gaussian variables and processes jointly by
$Z_{\lambda,k}=\bm b_{kk}^\top\mathbf\Sigma^{1/2}\bm G_n$ and
$\Xi_{k,n}(x)=\bm d_{k,J_\c}^\top(x)\mathbf\Sigma^{1/2}\bm G_n$.
Their variances are
$\Var(Z_{\lambda,k})=\sigma_{nk}^2$ and
$s_{\psi,k}^2(x):=\Var\{\Xi_{k,n}(x)\}
=\bm d_{k,J_\c}^\top(x)\mathbf\Sigma\bm d_{k,J_\c}(x)$,
which are exactly the asymptotic variances used in
Theorems~\ref{THM:eigenvalues} and~\ref{thm:fpcs}.

\begin{theorem}\label{thm:joint-eigensystem}
Under (S1)--(S4), the following conclusions hold.
\begin{itemize}
\item[(1)] For the eigenvalues,
\[
 \sup_{z\in\mathbb R}\left|\P\left(
 \max_{k\le K_n}\frac{\sqrt n|\widetilde\lambda_k-\lambda_k|}
 {\sigma_{nk}}\le z\right)
 -\P\left(\max_{k\le K_n}\frac{|Z_{\lambda,k}|}
 {\sigma_{nk}}\le z\right)\right|\to0.
\]
\item[(2)] For the eigenfunctions,
\[
 \begin{aligned}
 \sup_{z\in\mathbb R}\Bigg|\P\left(
 \max_{k\le K_n}\sup_{x\in[0,1]}
 \frac{\sqrt n|\widetilde\psi_k(x)-\psi_k(x)|}
 {s_{\psi,k}(x)}\le z\right)
 -\P\left(\max_{k\le K_n}\sup_{x\in[0,1]}
 \frac{|\Xi_{k,n}(x)|}{s_{\psi,k}(x)}\le z\right)\Bigg|\to0.
 \end{aligned}
\]
\item[(3)] Let $T_{\lambda,n}$ and $T_{\psi,n}$ be the empirical maxima
in (1) and (2), and let $T_{\lambda,n}^{G}$ and $T_{\psi,n}^{G}$ be
their Gaussian counterparts. Then
\[
 \sup_{z\in\mathbb R}\left|
 \P\{\max(T_{\lambda,n},T_{\psi,n})\le z\}
 -\P\{\max(T_{\lambda,n}^{G},T_{\psi,n}^{G})\le z\}
 \right|\to0.
\]
\end{itemize}
\end{theorem}

For fixed $0<\alpha<1$, the $(1-\alpha)$-quantile of the combined
Gaussian maximum in (3) gives intervals and bands that cover all the
first $K_n$ eigenvalues and eigenfunctions simultaneously with probability
$1-\alpha+o(1)$. For any fixed $k$, the same proof gives
$\sqrt n(\widetilde\lambda_k-\lambda_k)/\sigma_{nk}\Rightarrow N(0,1)$
and the approximation for the standardized eigenfunction supremum in
Theorem~\ref{thm:fpcs}, with the same Gaussian process $\Xi_{k,n}$.

% BLUE REVISION: Removed the Fourier-example paragraph and the paragraph
% announcing the two optional dependent-score extensions.

\subsection{Feasible covariance estimator}\label{sec:feasibleCov}

The feasible procedure first fits the mean by univariate B-spline least
squares and then forms covariance residual products from the same observations.

The two least-squares fits are
\begin{align*}
\widehat m
&=\mathop{\arg\min}_{g\in\mathcal S_{J_\m}^{p_\m}}
\sum_{i=1}^n\sum_{j=1}^{N_i}\{Y_{ij}-g(X_{ij})\}^2,
\qquad \widehat U_{ij}=Y_{ij}-\widehat m(X_{ij}),\\
\widehat G
&=\mathop{\arg\min}_{g\in
\mathcal S_{J_\c}^{p_\c}\otimes\mathcal S_{J_\c}^{p_\c}}
\sum_{i=1}^n\sum_{j_1\ne j_2}^{N_i}
\{g(X_{ij_1},X_{ij_2})-\widehat U_{ij_1}\widehat U_{ij_2}\}^2.
\end{align*}
Let $\widehat{\mathcal G}$ be the integral operator with kernel
$\{\widehat G(x,u)+\widehat G(u,x)\}/2$, with eigenpairs
$(\widehat\lambda_k,\widehat\psi_k)$.
Writing $\widehat G(x,x')=\bm B^\top(x,x')\widehat{\bm\theta}$, define
\(\widehat{\bm S}_i=\sum_{j_1\ne j_2}^{N_i} \bm B(X_{ij_1},X_{ij_2}) \{\widehat U_{ij_1}\widehat U_{ij_2} -\widehat G(X_{ij_1},X_{ij_2})\}.\)
The normal equations imply $\sum_i\widehat{\bm S}_i=\bm0$ exactly.
The subject-cluster sandwich is
\begin{equation}\label{eq:cluster-sandwich}
 \widehat{\mathbf\Sigma}=
 \frac1{nP_{n,2}^2(N)}\widehat{\mathbf Q}^{-1}
 \left(\sum_i\widehat{\bm S}_i\widehat{\bm S}_i^\top\right)
 \widehat{\mathbf Q}^{-1}.
\end{equation}

For the general oracle results in Sections~\ref{sec:infeasibleCov}--\ref{sec:eigenfunctions},
retain their standing design, count, moment, and spline conditions and add:
\begin{itemize}
\item[(F1)] $m\in\mathcal H_{q_\m,\nu_\m}[0,1]$, where $q_\m\ge1$ is
an integer, $\nu_\m\in(0,1]$, $q_\m^*=q_\m+\nu_\m>1$, and
$p_\m\ge q_\m^*$.
\item[(F2)] The mean and covariance splines have fixed orders and compatible
quasi-uniform partitions, $J_\m\asymp J_\c$, and
\[
\left\{\frac{J_\c^3(1+J_\c/\bar N)}n +J_\c^{2-2q_\m^*} +\frac{nJ_\c^{-4q_\m^*}}{(1+J_\c/\bar N)^2}\right\}\log n\to0.
\]
\end{itemize}
These restrictions control the mean approximation and stochastic estimation
errors.  

\begin{theorem}
\label{thm:feasible-same-sample}
Under the conditions of Theorem \ref{theorem1} and Assumptions
(F1)--(F2),
\[
 \sup_{x,x'\in[0,1]}
 \frac{\sqrt n|\widehat G(x,x')-\widetilde G(x,x')|}
 {\|\boldsymbol\Sigma^{1/2}\bm B(x,x')\|_2}
 =o_p\{(\log n)^{-1/2}\}.
\]
Thus, the oracle Gaussian approximation in Theorem \ref{theorem1}   
holds for $\widehat G$.
\end{theorem}

\begin{corollary}
\label{cor:feasible-fixed-eigensystem}
{Fix $K<\infty$ and assume (F1)--(F2) and}
\[
\sqrt{\log n}\left\{\frac{J_\c^3}{\sqrt n} +J_\c^{2-q_\m^*}+\sqrt nJ_\c^{-2q_\m^*}\right\}\to0.
\]
{Under the assumptions of Theorem~\ref{THM:eigenvalues},
its conclusion continues to hold for $\widehat\lambda_k$, $k\le K$.
Under the assumptions of Theorem~\ref{thm:fpcs},
its conclusion continues to hold for $\widehat\psi_k$, $k\le K$.} 
\end{corollary}

 For the growing collection in Section~\ref{sec:joint-eigensystem},
the corresponding feasible conclusion is as follows.

\begin{corollary}
\label{cor:feasible-growing-eigensystem}
Under (S1)--(S4), assume $\|m\|_{C^{p_\c}}\le C$ and fit the mean in
$\mathcal S_{J_\c}^{p_\c}$. Then, the conclusions of
Theorem~\ref{thm:joint-eigensystem} hold with
$(\widetilde\lambda_k,\widetilde\psi_k)$ replaced by
$(\widehat\lambda_k,\widehat\psi_k)$.
\end{corollary}
The conditions of this corollary imply (F1)--(F2), with
$q_\m=p_\c-1$, $\nu_\m=1$, $p_\m=p_\c$, and $J_\m=J_\c$.

{\section{Implementation}\label{sec:implement}
%In the computations, we multiply the sandwich estimator in \eqref{eq:cluster-sandwich} by $n/(n-1)$.

For numerical stability, after computing $\widehat m$ and $\widehat G$ by the least-squares
fits in Section~\ref{sec:feasibleCov}, 
one symmetrizes the fitted covariance coefficient matrix
and replaces the sandwich estimator $\widehat{\boldsymbol\Sigma}$ in
\eqref{eq:cluster-sandwich} by the positive semidefinite
part of its symmetrization, i.e., truncating negative eigenvalues
due to numerical roundoff to zero.

\subsection{SCR for covariance}

Let $\mathcal X_{\mathrm{ev}}\subset[0,1]$ be the evaluation grid for the
confidence bands. Using the basis $\bm B(x)$ and tensor basis
$\bm B(x,x')$ from Section~\ref{sec:infeasibleCov}, generate
$B_{\mathrm{mc}}$ independent copies $\bm G_n^{(b)}$ of the standard
Gaussian vector in Theorem~\ref{theorem1}. The corresponding fitted
Gaussian surfaces are
\(
 \widehat{\mathcal Z}_n^{(b)}(x,x')
 =\bm B^\top(x,x')\widehat{\mathbf\Sigma}^{1/2}\bm G_n^{(b)},
 \ b=1,\ldots,B_{\mathrm{mc}}.
\)
For each draw, divide this surface by
$\|\widehat{\mathbf\Sigma}^{1/2}\bm B(x,x')\|_2$ and take its maximum
absolute value over $(x,x')\in\mathcal X_{\mathrm{ev}}^2$.
The empirical $(1-\alpha)$ quantile of these maxima gives the covariance
SCR critical value. Center the SCR at $\widehat G(x,x')$, with half-width
equal to this critical value times
$\|\widehat{\mathbf\Sigma}^{1/2}\bm B(x,x')\|_2/\sqrt n$.

\subsection{Gaussian approximation for eigensystem inference}\label{sec:asymp-calibration}

For eigensystem inference, let $\widehat{\boldsymbol\Theta}$ be the
symmetric coefficient matrix corresponding to $\widehat{\bm\theta}$ in
Section~\ref{sec:feasibleCov}. With column-stacking vectorization,
$\vec(\widehat{\boldsymbol\Theta}^{\top})=\widehat{\bm\theta}$, so
$\widehat G(x,x')=\bm B^\top(x)\widehat{\boldsymbol\Theta}\bm B(x')$.
On integration nodes $x_1,\ldots,x_M$, form
$\mathbf B_M=\{\bm B^\top(x_j)\}_{j=1}^M$ and use the integration matrix
$\mathbf V_M=M^{-1}\mathbf B_M^\top\mathbf B_M$ to approximate
$\int_0^1\bm B(x)\bm B^\top(x)\,dx$.
Take its Cholesky decomposition $\mathbf V_M=\mathbf L_M\mathbf L_M^\top$.
Let $\widehat\lambda_{k,M}$ and $\widehat{\bm v}_{k,M}$ be the eigenvalues
and unit eigenvectors of
$\mathbf L_M^\top\widehat{\boldsymbol\Theta}\mathbf L_M$. Set 
\(\widehat{\bm\gamma}_{k,M}=(\mathbf L_M^\top)^{-1}\widehat{\bm v}_{k,M}, \ \widehat\psi_{k,M}(x)=\bm B^\top(x)\widehat{\bm\gamma}_{k,M}, \)
\[
\widehat{\bm u}_{k,M}=\mathbf V_M\widehat{\bm\gamma}_{k,M} =M^{-1}\sum_{j=1}^M\bm B(x_j)\widehat\psi_{k,M}(x_j).
\]
The pairs $(\widehat\lambda_{k,M},\widehat\psi_{k,M})$ numerically
approximate $(\widehat\lambda_k,\widehat\psi_k)$ from
Section~\ref{sec:feasibleCov}; their eigenfunctions have unit norm in the
equal-weight inner product on the integration nodes.
The eigenproblem has dimension $J_\c+p_\c$, so $M$ can increase without
increasing this dimension. Retain the positive eigenvalues of interest and evaluate their eigenfunctions
on $\mathcal X_{\mathrm{ev}}$, fixing signs using the anchors in (S3)
and inner products on the integration nodes.

We denote this procedure by Asymp.
The numerical plug-in versions of $\bm b_{\ell k}$ and the symmetrized
$\bm d_{k,J_\c}(x)$ from Sections~\ref{sec:FPCA}--\ref{sec:eigenfunctions}
are
\begin{align*}
 \widehat{\bm b}_{\ell k,M}
 &=\widehat{\bm u}_{\ell,M}\otimes\widehat{\bm u}_{k,M},\quad
 \widehat{\bm d}_{k,J_\c,M}(x)
 =\sum_{\substack{1\le\ell\le J_\c+p_\c\\\ell\ne k}}
 \frac{\widehat\psi_{\ell,M}(x)}
      {\widehat\lambda_{k,M}-\widehat\lambda_{\ell,M}}
 \frac{\widehat{\bm b}_{\ell k,M}+\widehat{\bm b}_{k\ell,M}}{2}.
\end{align*}
The symmetrization accounts for symmetric covariance perturbations.
The sum uses the full fitted spectrum, including zero and negative
eigenvalues. Applying these vectors to the common Gaussian draws gives
the fitted counterparts of $Z_{\lambda,k}$ and $\Xi_{k,n}(x)$ in
Section~\ref{sec:joint-eigensystem}:
\(
\widehat Z_{\lambda,k,M}^{(b)}
  =\widehat{\bm b}_{kk,M}^\top\widehat{\mathbf\Sigma}^{1/2}\bm G_n^{(b)}\) and \(\widehat\Xi_{k,n,M}^{(b)}(x)
  =\widehat{\bm d}_{k,J_\c,M}^\top(x)\widehat{\mathbf\Sigma}^{1/2}\bm G_n^{(b)}.
\)
Their conditional variances are
\(
\widehat\sigma_{nk,M}^2
  =\widehat{\bm b}_{kk,M}^\top\widehat{\mathbf\Sigma}\widehat{\bm b}_{kk,M}\) and \(\widehat s_{\psi,k,M}^2(x)
  =\widehat{\bm d}_{k,J_\c,M}^\top(x)\widehat{\mathbf\Sigma}
   \widehat{\bm d}_{k,J_\c,M}(x).
\)
Standardize the simulated values by $\widehat\sigma_{nk,M}$ and
$\widehat s_{\psi,k,M}(x)$, respectively,
and take the maximum absolute value over the selected eigenvalues, over all
selected eigenfunction--grid pairs with $x\in\mathcal X_{\mathrm{ev}}$,
or over both families jointly.
For joint inference in Section~\ref{sec:joint-eigensystem}, the selected
indices are $1\leq k\leq K_n$.
The corresponding empirical $(1-\alpha)$ quantiles give the critical values.
Center the intervals and bands at $\widehat\lambda_{k,M}$ and
$\widehat\psi_{k,M}(x)$, with half-widths equal to the corresponding critical
value times $\widehat\sigma_{nk,M}/\sqrt n$ and
$\widehat s_{\psi,k,M}(x)/\sqrt n$, respectively.
For inference on an individual component, restrict the maximum to that
component. For an individual eigenvalue, use the standard-normal critical
value $z_{1-\alpha/2}$. Using common Gaussian draws preserves the
dependence across components and families. The fitted
eigensystem and derivative vectors are computed once and reused
throughout the Gaussian simulation.

\subsection{Direct Gaussian perturbation for eigensystem inference}\label{sec:perturb-calibration}
As illustrated in Section~\ref{sec:simulation}, the first-order approximation can give inadequate eigenfunction coverage
in small samples. Therefore,
we propose a direct Gaussian perturbation of the fitted
covariance, denoted by Perturb. This procedure recomputes the eigensystem
after each perturbation and thus retains its nonlinear response to
covariance estimation error.

For each Gaussian draw, form the symmetric spline kernel
\[
 \widehat G^{*(b)}(x,x')=\widehat G(x,x')
 +\frac{\widehat{\mathcal Z}_n^{(b)}(x,x')+
        \widehat{\mathcal Z}_n^{(b)}(x',x)}{2\sqrt n}
\]
and compute its numerical eigenpairs
$(\widehat\lambda_{k,M}^{*(b)},\widehat\psi_{k,M}^{*(b)})$ using the same
integration matrix $\mathbf V_M$.
Order the perturbed eigenvalues and align each eigenfunction with
$\widehat\psi_{k,M}$ using the equal-weight inner product on the
integration nodes. Replace $\widehat Z_{\lambda,k,M}^{(b)}$ and
$\widehat\Xi_{k,n,M}^{(b)}(x)$ by $\sqrt n\{\widehat\lambda_{k,M}^{*(b)}-\widehat\lambda_{k,M}\}$ and $\sqrt n\{\widehat\psi_{k,M}^{*(b)}(x)-\widehat\psi_{k,M}(x)\}.$
Using the original fitted scales $\widehat\sigma_{nk,M}$ and
$\widehat s_{\psi,k,M}(x)$, standardize these differences
and compute the same absolute maxima and empirical quantiles as in
Section~\ref{sec:asymp-calibration}. For an individual eigenvalue, the critical
value is the empirical quantile of its absolute standardized perturbation. Here, we note that
the two procedures use the same point estimates and standard errors,
with different critical values.}

\section{Simulation study}\label{sec:simulation}
\subsection{Simulation setup}\label{sec:simulation-setup}
We study the accuracy of covariance and eigensystem inference under different sampling frequencies, using the proposed procedures in Section~\ref{sec:implement}.

Data are generated from \eqref{eq:observation-model} with four components,
$m(t)=2\sin(\pi t)$, $\lambda_k=2^{-k}$ and
$\psi_k(t)=\sqrt{2k-1}\,P_{k-1}(2t-1)$, $k=1,\ldots,4$, where $P_d$ is
the Legendre polynomial of degree $d$. We generate $\xi_{ik}\sim N(0,1)$,
$\epsilon_{ij}\sim N(0,1)$ and $X_{ij}\sim U[0,1]$ independently, and set
$\sigma^2(x)\equiv0.1$.
We take $n=400$, 800, 1600, 3200 and 6400. Independently of the
trajectories, errors and design points, the counts $N_i$ are uniform on the
integers in one of four ranges: $[6,9]$, $[\lceil2n^{1/3}\rceil,\lfloor3n^{1/3}\rfloor]$, $[\lceil2n^{1/2}\rceil,\lfloor3n^{1/2}\rfloor]$ and $[\lfloor0.6n\rfloor,\lfloor0.8n\rfloor]$.
These define Regimes 1--4, respectively.

Both the mean and covariance fits use cubic B-splines of order
$p_\m=p_\c=4$, with the same equally spaced internal knots. We take $J_\m=J_\c=\left\lceil Cn^{2/15}\right\rceil.$
The main experiments fix $C=1$. The supplement examines sensitivity to
$C$ by reporting additional results for $C\in\{0.5,0.75,1.5,2\}$.
We may take $q^\ast=q_\m^\ast=4$ and $r=16$. The choice
$J_\c\asymp n^{2/15}$ satisfies the sufficient dimension conditions in
(A7), (B2), (C2), and (F2) across the four sampling regimes.

For spectral inference, we compare Asymp. and Perturb. procedures defined in
Sections~\ref{sec:asymp-calibration} and~\ref{sec:perturb-calibration} respectively.
We fix the signs using the deterministic anchors $a_k=\psi_k$.

We use 500 Monte Carlo replications per setting and $B_{\mathrm{mc}}=500$ Gaussian draws
to estimate critical values for the nominal $95\%$ confidence sets. {For numerical integration, take $M=10n$ and $x_j=j/M$, $j=1,\ldots,M$.
These nodes give
$\mathbf V_M=M^{-1}\sum_{j=1}^M\bm B(x_j)\bm B^\top(x_j)$ for the
coefficient-space eigendecomposition in Section~\ref{sec:implement}.
For evaluation, use the separate grid
$\mathcal X_{\mathrm{ev}}=\{(j-1)/49:j=1,\ldots,50\}$, including both
endpoints. Evaluate $\widehat G$ on $\mathcal X_{\mathrm{ev}}^2$ and
$\widehat\psi_{k,M}$ on $\mathcal X_{\mathrm{ev}}$.
The corresponding Gaussian surfaces and eigenfunction draws are evaluated
on the same grids to obtain the SCR critical values. Coverage and average
widths are also computed on these grids. Both spectral inference procedures use the same simulated data, point estimates, integration
nodes and evaluation grid.}

We examine three inferential targets: the covariance function SCR; the eigenfunction SCR and the eigenvalue CI of each fixed
component $k=1,2,3,4$; and the joint eigenfunction SCR and the joint eigenvalue CI families for $1\leq k\leq K_n$, where
$K_n=\lfloor0.5\log n\rfloor$.  

\subsection{Covariance SCRs}
Table~\ref{tab:organized-covariance} reports covariance coverage and average
SCR width for the four sampling regimes. It shows that covariance SCRs narrow as the sample size and sampling frequency increase.
Also coverage approaches the nominal level with increasing $n$.

\begin{table}[htbp]
\centering
\small\linespread{1}\selectfont
\caption{Coverage probability and average full width (in parentheses) of nominal $95\%$ covariance SCRs. }
\label{tab:organized-covariance}
\small\setlength{\tabcolsep}{5pt}\renewcommand{\arraystretch}{1.02}
\begin{tabular}{rrrrr}
\toprule
$n$ & Regime 1 & Regime 2 & Regime 3 & Regime 4 \\
\midrule
400 & 0.720 (0.770) & 0.826 (0.520) & 0.866 (0.410) & 0.924 (0.343) \\
800 & 0.834 (0.554) & 0.882 (0.348) & 0.924 (0.277) & 0.908 (0.238) \\
1600 & 0.890 (0.401) & 0.916 (0.232) & 0.926 (0.189) & 0.944 (0.167) \\
3200 & 0.914 (0.285) & 0.928 (0.156) & 0.944 (0.129) & 0.944 (0.117) \\
6400 & 0.922 (0.224) & 0.952 (0.110) & 0.940 (0.091) & 0.946 (0.083) \\
\bottomrule
\end{tabular}
\end{table}

\subsection{Inference for fixed components}
{Eigenvalue coverage and CI widths are similar for the two methods (Table~\ref{tab:organized-fixed-lambda}).
For higher components in Regime~1, Perturb. gives wider eigenfunction SCRs and higher coverage in small samples.
These differences are smaller at the largest sample size (Table~\ref{tab:organized-fixed-psi}).
For $k=4$ and $n=400$ in Regime~1, the mean additional time to construct the eigenvalue CI and eigenfunction SCR was 1.8 ms for Asymp. and 83.4 ms for Perturb. on the same machine.}
\begin{table}[!htbp]
\centering
\small\linespread{1}\selectfont
\caption{Coverage probability and average full width (in parentheses) of nominal $95\%$ eigenvalue CIs for each fixed $k$.}
\label{tab:organized-fixed-lambda}
\footnotesize\setlength{\tabcolsep}{5pt}\renewcommand{\arraystretch}{1}
\begin{tabular}{rrlrrrr}
\toprule
$k$ & $n$ & Method & Regime 1 & Regime 2 & Regime 3 & Regime 4 \\
\midrule
1 & 400 & Asymp. & 0.964 (0.169) & 0.940 (0.151) & 0.944 (0.145) & 0.954 (0.140) \\
 &  & Perturb. & 0.968 (0.170) & 0.942 (0.151) & 0.940 (0.145) & 0.948 (0.140) \\
 & 800 & Asymp. & 0.952 (0.117) & 0.962 (0.105) & 0.952 (0.102) & 0.944 (0.099) \\
 &  & Perturb. & 0.952 (0.118) & 0.960 (0.105) & 0.950 (0.102) & 0.940 (0.099) \\
 & 1600 & Asymp. & 0.950 (0.084) & 0.946 (0.074) & 0.966 (0.072) & 0.944 (0.070) \\
 &  & Perturb. & 0.954 (0.084) & 0.946 (0.074) & 0.960 (0.072) & 0.940 (0.070) \\
 & 3200 & Asymp. & 0.952 (0.059) & 0.934 (0.052) & 0.932 (0.051) & 0.934 (0.050) \\
 &  & Perturb. & 0.944 (0.059) & 0.942 (0.052) & 0.938 (0.051) & 0.936 (0.050) \\
 & 6400 & Asymp. & 0.934 (0.041) & 0.952 (0.036) & 0.954 (0.036) & 0.946 (0.035) \\
 &  & Perturb. & 0.930 (0.042) & 0.950 (0.036) & 0.954 (0.036) & 0.950 (0.035) \\
\midrule
2 & 400 & Asymp. & 0.946 (0.122) & 0.938 (0.090) & 0.940 (0.078) & 0.938 (0.070) \\
 &  & Perturb. & 0.940 (0.119) & 0.936 (0.089) & 0.942 (0.077) & 0.940 (0.070) \\
 & 800 & Asymp. & 0.950 (0.086) & 0.948 (0.061) & 0.926 (0.053) & 0.930 (0.050) \\
 &  & Perturb. & 0.948 (0.085) & 0.942 (0.061) & 0.932 (0.053) & 0.920 (0.050) \\
 & 1600 & Asymp. & 0.960 (0.061) & 0.946 (0.042) & 0.942 (0.037) & 0.950 (0.035) \\
 &  & Perturb. & 0.960 (0.061) & 0.948 (0.042) & 0.942 (0.037) & 0.944 (0.035) \\
 & 3200 & Asymp. & 0.938 (0.043) & 0.944 (0.028) & 0.966 (0.026) & 0.940 (0.025) \\
 &  & Perturb. & 0.938 (0.043) & 0.944 (0.028) & 0.964 (0.026) & 0.938 (0.025) \\
 & 6400 & Asymp. & 0.946 (0.030) & 0.950 (0.020) & 0.962 (0.018) & 0.950 (0.018) \\
 &  & Perturb. & 0.950 (0.030) & 0.954 (0.020) & 0.962 (0.018) & 0.944 (0.018) \\
\midrule
3 & 400 & Asymp. & 0.956 (0.090) & 0.940 (0.056) & 0.938 (0.042) & 0.920 (0.036) \\
 &  & Perturb. & 0.946 (0.084) & 0.932 (0.055) & 0.934 (0.042) & 0.912 (0.036) \\
 & 800 & Asymp. & 0.948 (0.063) & 0.940 (0.037) & 0.942 (0.029) & 0.944 (0.025) \\
 &  & Perturb. & 0.936 (0.061) & 0.934 (0.037) & 0.950 (0.029) & 0.946 (0.025) \\
 & 1600 & Asymp. & 0.954 (0.046) & 0.938 (0.024) & 0.952 (0.020) & 0.952 (0.018) \\
 &  & Perturb. & 0.948 (0.045) & 0.936 (0.024) & 0.952 (0.020) & 0.954 (0.018) \\
 & 3200 & Asymp. & 0.950 (0.032) & 0.914 (0.016) & 0.950 (0.013) & 0.948 (0.012) \\
 &  & Perturb. & 0.948 (0.032) & 0.916 (0.016) & 0.948 (0.013) & 0.952 (0.012) \\
 & 6400 & Asymp. & 0.956 (0.023) & 0.956 (0.011) & 0.948 (0.009) & 0.960 (0.009) \\
 &  & Perturb. & 0.958 (0.023) & 0.954 (0.011) & 0.948 (0.009) & 0.958 (0.009) \\
\midrule
4 & 400 & Asymp. & 0.932 (0.068) & 0.912 (0.037) & 0.936 (0.025) & 0.948 (0.019) \\
 &  & Perturb. & 0.924 (0.062) & 0.910 (0.037) & 0.938 (0.025) & 0.948 (0.019) \\
 & 800 & Asymp. & 0.932 (0.050) & 0.958 (0.024) & 0.942 (0.016) & 0.934 (0.013) \\
 &  & Perturb. & 0.930 (0.048) & 0.954 (0.024) & 0.940 (0.016) & 0.934 (0.013) \\
 & 1600 & Asymp. & 0.952 (0.037) & 0.948 (0.015) & 0.946 (0.011) & 0.934 (0.009) \\
 &  & Perturb. & 0.946 (0.036) & 0.950 (0.015) & 0.944 (0.011) & 0.930 (0.009) \\
 & 3200 & Asymp. & 0.936 (0.026) & 0.944 (0.010) & 0.944 (0.007) & 0.946 (0.006) \\
 &  & Perturb. & 0.934 (0.026) & 0.944 (0.010) & 0.948 (0.007) & 0.940 (0.006) \\
 & 6400 & Asymp. & 0.954 (0.019) & 0.944 (0.007) & 0.942 (0.005) & 0.946 (0.004) \\
 &  & Perturb. & 0.948 (0.019) & 0.944 (0.007) & 0.946 (0.005) & 0.952 (0.004) \\
\bottomrule
\end{tabular}
\end{table}
\begin{table}[!htbp]
\centering
\small\linespread{1}\selectfont
\caption{Coverage probability and average full width (in parentheses) of nominal $95\%$ eigenfunction SCRs for each fixed $k$.}
\label{tab:organized-fixed-psi}
\footnotesize\setlength{\tabcolsep}{5pt}\renewcommand{\arraystretch}{1}
\begin{tabular}{rrlrrrr}
\toprule
$k$ & $n$ & Method & Regime 1 & Regime 2 & Regime 3 & Regime 4 \\
\midrule
1 & 400 & Asymp. & 0.922 (0.792) & 0.898 (0.583) & 0.926 (0.477) & 0.914 (0.416) \\
 &  & Perturb. & 0.930 (0.822) & 0.906 (0.600) & 0.938 (0.494) & 0.930 (0.433) \\
 & 800 & Asymp. & 0.956 (0.561) & 0.920 (0.389) & 0.936 (0.323) & 0.948 (0.288) \\
 &  & Perturb. & 0.960 (0.568) & 0.928 (0.394) & 0.946 (0.329) & 0.956 (0.293) \\
 & 1600 & Asymp. & 0.944 (0.395) & 0.930 (0.261) & 0.938 (0.222) & 0.956 (0.200) \\
 &  & Perturb. & 0.942 (0.397) & 0.936 (0.263) & 0.940 (0.223) & 0.954 (0.202) \\
 & 3200 & Asymp. & 0.920 (0.280) & 0.948 (0.177) & 0.948 (0.152) & 0.940 (0.141) \\
 &  & Perturb. & 0.920 (0.280) & 0.946 (0.178) & 0.948 (0.153) & 0.938 (0.142) \\
 & 6400 & Asymp. & 0.936 (0.208) & 0.956 (0.124) & 0.934 (0.107) & 0.944 (0.099) \\
 &  & Perturb. & 0.934 (0.208) & 0.956 (0.124) & 0.932 (0.107) & 0.944 (0.100) \\
\midrule
2 & 400 & Asymp. & 0.886 (1.349) & 0.942 (0.892) & 0.932 (0.692) & 0.932 (0.588) \\
 &  & Perturb. & 0.930 (1.581) & 0.956 (0.943) & 0.956 (0.726) & 0.946 (0.618) \\
 & 800 & Asymp. & 0.912 (0.944) & 0.910 (0.590) & 0.956 (0.464) & 0.934 (0.400) \\
 &  & Perturb. & 0.936 (0.996) & 0.918 (0.601) & 0.962 (0.473) & 0.944 (0.409) \\
 & 1600 & Asymp. & 0.936 (0.667) & 0.944 (0.388) & 0.934 (0.314) & 0.956 (0.279) \\
 &  & Perturb. & 0.942 (0.682) & 0.944 (0.391) & 0.940 (0.317) & 0.964 (0.283) \\
 & 3200 & Asymp. & 0.950 (0.472) & 0.962 (0.260) & 0.946 (0.215) & 0.944 (0.196) \\
 &  & Perturb. & 0.956 (0.476) & 0.962 (0.261) & 0.950 (0.216) & 0.948 (0.197) \\
 & 6400 & Asymp. & 0.938 (0.352) & 0.952 (0.179) & 0.944 (0.150) & 0.948 (0.138) \\
 &  & Perturb. & 0.938 (0.352) & 0.952 (0.180) & 0.946 (0.151) & 0.950 (0.138) \\
\midrule
3 & 400 & Asymp. & 0.802 (1.996) & 0.900 (1.143) & 0.926 (0.778) & 0.954 (0.606) \\
 &  & Perturb. & 0.952 (3.111) & 0.938 (1.336) & 0.958 (0.837) & 0.968 (0.646) \\
 & 800 & Asymp. & 0.856 (1.408) & 0.924 (0.716) & 0.948 (0.507) & 0.946 (0.405) \\
 &  & Perturb. & 0.944 (1.793) & 0.936 (0.747) & 0.958 (0.522) & 0.952 (0.417) \\
 & 1600 & Asymp. & 0.898 (1.003) & 0.934 (0.460) & 0.940 (0.337) & 0.954 (0.280) \\
 &  & Perturb. & 0.932 (1.094) & 0.940 (0.467) & 0.950 (0.342) & 0.960 (0.284) \\
 & 3200 & Asymp. & 0.932 (0.696) & 0.944 (0.302) & 0.964 (0.226) & 0.948 (0.196) \\
 &  & Perturb. & 0.946 (0.716) & 0.946 (0.304) & 0.966 (0.227) & 0.952 (0.197) \\
 & 6400 & Asymp. & 0.938 (0.522) & 0.956 (0.205) & 0.946 (0.156) & 0.942 (0.138) \\
 &  & Perturb. & 0.940 (0.528) & 0.954 (0.205) & 0.950 (0.157) & 0.944 (0.138) \\
\midrule
4 & 400 & Asymp. & 0.762 (2.635) & 0.886 (1.235) & 0.930 (0.726) & 0.932 (0.479) \\
 &  & Perturb. & 0.964 (4.504) & 0.934 (1.443) & 0.952 (0.782) & 0.950 (0.509) \\
 & 800 & Asymp. & 0.852 (1.818) & 0.944 (0.756) & 0.932 (0.455) & 0.944 (0.314) \\
 &  & Perturb. & 0.940 (2.491) & 0.948 (0.790) & 0.950 (0.469) & 0.950 (0.323) \\
 & 1600 & Asymp. & 0.884 (1.253) & 0.916 (0.466) & 0.956 (0.292) & 0.944 (0.212) \\
 &  & Perturb. & 0.928 (1.418) & 0.930 (0.475) & 0.958 (0.296) & 0.950 (0.215) \\
 & 3200 & Asymp. & 0.946 (0.882) & 0.950 (0.296) & 0.934 (0.189) & 0.948 (0.147) \\
 &  & Perturb. & 0.954 (0.916) & 0.956 (0.299) & 0.938 (0.190) & 0.952 (0.148) \\
 & 6400 & Asymp. & 0.944 (0.687) & 0.954 (0.202) & 0.950 (0.129) & 0.954 (0.103) \\
 &  & Perturb. & 0.950 (0.699) & 0.956 (0.202) & 0.952 (0.130) & 0.958 (0.104) \\
\bottomrule
\end{tabular}
\end{table}

\subsection{Joint eigensystem inference}
{Table~\ref{tab:organized-joint} gives joint coverage and average widths for $k\le K_n$, using separate critical values for the eigenvalue and eigenfunction families.}
\begin{table}[ht]
\centering
\small\linespread{1}\selectfont
\caption{Joint coverage probability and average full width (in parentheses) of nominal $95\%$ eigenvalue CIs and eigenfunction SCRs.}
\label{tab:organized-joint}
\small\setlength{\tabcolsep}{5pt}\renewcommand{\arraystretch}{1.02}
\begin{tabular}{rrlrrrr}
\toprule
$n$ & $K_n$ & Method & Regime 1 & Regime 2 & Regime 3 & Regime 4 \\
\midrule\multicolumn{7}{c}{Joint eigenvalue CIs} \\
\midrule
400 & 2 & Asymp. & 0.954 (0.165) & 0.928 (0.138) & 0.940 (0.127) & 0.936 (0.120) \\
 &  & Perturb. & 0.948 (0.165) & 0.928 (0.136) & 0.942 (0.127) & 0.934 (0.120) \\
800 & 3 & Asymp. & 0.932 (0.108) & 0.940 (0.082) & 0.928 (0.075) & 0.936 (0.071) \\
 &  & Perturb. & 0.932 (0.107) & 0.938 (0.082) & 0.926 (0.075) & 0.930 (0.071) \\
1600 & 3 & Asymp. & 0.950 (0.077) & 0.928 (0.057) & 0.934 (0.052) & 0.944 (0.050) \\
 &  & Perturb. & 0.948 (0.077) & 0.926 (0.057) & 0.930 (0.052) & 0.944 (0.050) \\
3200 & 4 & Asymp. & 0.934 (0.051) & 0.928 (0.034) & 0.944 (0.031) & 0.952 (0.030) \\
 &  & Perturb. & 0.934 (0.050) & 0.932 (0.034) & 0.944 (0.031) & 0.952 (0.030) \\
6400 & 4 & Asymp. & 0.950 (0.036) & 0.956 (0.023) & 0.956 (0.022) & 0.952 (0.021) \\
 &  & Perturb. & 0.954 (0.036) & 0.956 (0.023) & 0.958 (0.022) & 0.950 (0.021) \\
\midrule\multicolumn{7}{c}{Joint eigenfunction SCRs} \\
\midrule
400 & 2 & Asymp. & 0.870 (1.151) & 0.896 (0.792) & 0.920 (0.628) & 0.926 (0.540) \\
 &  & Perturb. & 0.928 (1.313) & 0.928 (0.835) & 0.938 (0.658) & 0.946 (0.567) \\
800 & 3 & Asymp. & 0.864 (1.089) & 0.916 (0.632) & 0.928 (0.484) & 0.940 (0.408) \\
 &  & Perturb. & 0.952 (1.297) & 0.944 (0.657) & 0.954 (0.499) & 0.958 (0.421) \\
1600 & 3 & Asymp. & 0.900 (0.770) & 0.938 (0.414) & 0.930 (0.326) & 0.962 (0.283) \\
 &  & Perturb. & 0.936 (0.818) & 0.940 (0.421) & 0.940 (0.331) & 0.964 (0.288) \\
3200 & 4 & Asymp. & 0.930 (0.666) & 0.934 (0.298) & 0.956 (0.226) & 0.926 (0.196) \\
 &  & Perturb. & 0.946 (0.689) & 0.942 (0.300) & 0.960 (0.228) & 0.930 (0.198) \\
6400 & 4 & Asymp. & 0.930 (0.505) & 0.962 (0.203) & 0.946 (0.156) & 0.958 (0.138) \\
 &  & Perturb. & 0.934 (0.512) & 0.962 (0.204) & 0.954 (0.157) & 0.960 (0.139) \\
\bottomrule
\end{tabular}
\end{table}
Joint eigenvalue coverage and CI widths are similar for the two methods.
For eigenfunctions, Perturb. improves joint coverage in the smaller samples
and gives wider SCRs. In  large samples, the two methods give similar joint
coverage and widths.

\FloatBarrier

\section{Applications}\label{sec:application}
\newcommand{\PBCSubjects}{285}
\newcommand{\PBCMinCount}{2}
\newcommand{\PBCMedianCount}{6}
\newcommand{\PBCMaxCount}{10}
\newcommand{\PBCFirstEigenvalue}{0.972}
\newcommand{\PBCFirstEigenvalueLower}{0.773}
\newcommand{\PBCFirstEigenvalueUpper}{1.172}
\newcommand{\PBCFirstSpectrumPercent}{57.8}
\newcommand{\PBCCovPositivePercent}{94.3}
\newcommand{\TecatorSubjects}{193}
\newcommand{\TecatorObservations}{100}
\newcommand{\TecatorFirstEigenvalue}{0.257}
\newcommand{\TecatorFirstEigenvalueLower}{0.194}
\newcommand{\TecatorFirstEigenvalueUpper}{0.319}
\newcommand{\TecatorFirstSpectrumPercent}{98.7}
\newcommand{\TecatorPeakWavelength}{998}
\newcommand{\TecatorFatCorrelation}{0.427}
\newcommand{\TecatorMoistureCorrelation}{-0.459}
\newcommand{\TecatorProteinCorrelation}{-0.281}
{We apply the method to sparse longitudinal bilirubin measurements and densely sampled near-infrared spectra.} Both analyses use cubic B-splines and the subject-cluster sandwich in
Section~\ref{sec:implement}, with the Asymp. procedure in
Section~\ref{sec:asymp-calibration} for eigensystem inference.
We map each observation domain to $[0,1]$. Eigenvalues and eigenfunctions follow this normalization, while
figures display years or wavelengths. The eigenfunction signs are chosen
against the constant positive anchor. We use 2000 Gaussian coefficient draws to estimate separate critical values
for the covariance and first-eigenfunction bands. All intervals and bands have nominal level $95\%$. We report
the leading share of the fitted positive spectrum
$\widehat\lambda_1/\sum_{\widehat\lambda_r>0}\widehat\lambda_r$.

\subsection{Sparse functional data: bilirubin profiles}
Serial bilirubin measurements provide information about the clinical
course of primary biliary cirrhosis \citep{Murtaugh1994PBC}. We examine
whether differences in log-bilirubin levels persist across follow-up and
which periods receive the greatest weight in the leading mode of
between-patient variation.

We analyze the first eight years of follow-up using \texttt{pbcseq} from
the R package \texttt{survival} and retain positive bilirubin measurements,
average same-day measurements within each patient on the original scale,
and take natural logarithms. Including patients with at least two distinct
observation days gives \PBCSubjects{} patients and 1736 observations, with
\PBCMinCount{}--\PBCMaxCount{} measurements per patient and a median of
\PBCMedianCount{}. Bilirubin is measured in mg/dl, and days since enrollment
are divided by $8\times365.25$ for fitting. Both the mean and covariance
fits use four equally spaced internal knots.

\begin{figure}[!htbp]
\centering
\includegraphics[width=0.96\textwidth]{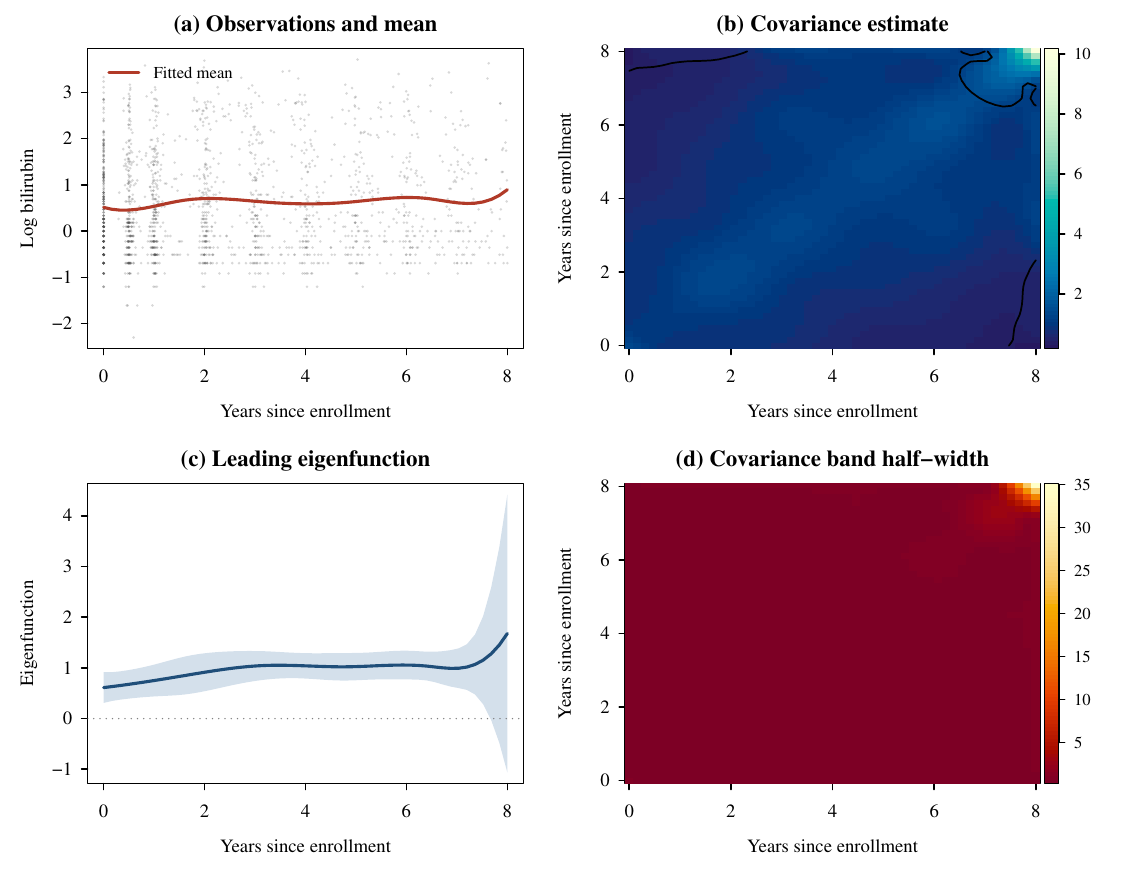}
\caption{Sparse longitudinal log-bilirubin data. (a) Observations and fitted
mean. (b) Estimated covariance; solid and dashed contours, where present,
mark zero levels of the lower and upper simultaneous limits.
(c) Leading eigenfunction and its simultaneous band.
(d) Half-width of the covariance band. Covariance and band half-width are
on the squared log-bilirubin scale.}
\label{fig:application-pbc}
\end{figure}

The fitted covariance supports persistent differences in log-bilirubin
levels across follow-up: its lower band is positive at
\PBCCovPositivePercent\% of the evaluation-grid pairs
(Figure~\ref{fig:application-pbc}). The leading eigenfunction describes
departures from the mean in the same direction over much of follow-up.
The leading component accounts for \PBCFirstSpectrumPercent\% of the fitted
positive spectrum; its eigenvalue is \PBCFirstEigenvalue{}, with marginal interval
$[\PBCFirstEigenvalueLower{},\PBCFirstEigenvalueUpper{}]$.
The largest estimated absolute loading occurs at eight years, where the
band widens markedly.

Using three to five internal knots in both fits gives first eigenvalues
between $0.938$ and $0.979$, with the largest absolute loading remaining
at eight years, and the leading share of the positive spectrum  ranges from $55.7\%$ to $67.3\%$. {Taken together, the predominantly positive covariance
and the same-sign leading eigenfunction suggest that
between-patient differences in log-bilirubin levels
persist throughout follow-up, rather than reflecting
contrasts between early and late periods.
Besides, the leading component captures a substantial proportion
of this variation, while the remaining components
reflect additional patterns of between-patient
heterogeneity.}
% {Some patients have fewer than four visits, and additional visits may be
% prompted by worsening disease. Follow-up and missingness near death or
% transplantation can be informative, as documented for \texttt{pbcseq}.\footnote{\url{https://stat.ethz.ch/R-manual/R-devel/library/survival/html/pbcseq.html}}
% We interpret the intervals and bands as descriptive uncertainty summaries
% under this observation process.}

\subsection{Dense functional data: near-infrared spectra}
{Tecator supplied meat absorbance spectra recorded with an Infratec Food and Feed Analyzer \citep{BorggaardThodberg1992,TecatorData}.} Each spectrum has 100
measurements from 850 to 1050~nm, accompanied by measurements of fat,
moisture and protein. Meat absorbance reflects the combined optical
contributions of these constituents \citep{ElMasryNakauchi2016}.
We examine whether absorbance varies through a common mode across
wavelengths and how this mode is associated with the recorded composition.

We pool the first 215 records, comprising the training,
monitoring and test sets for interpolation studies, and exclude its 25
extrapolation records. Collapsing 22 exact duplicate records with identical
spectra and composition measurements leaves \TecatorSubjects{} distinct
curves. We use the absorbance
measurements, map wavelength $w$ to $(w-850)/200$ and use five equally
spaced internal knots in both spline fits.

\begin{figure}[!htbp]
\centering
\includegraphics[width=0.96\textwidth]{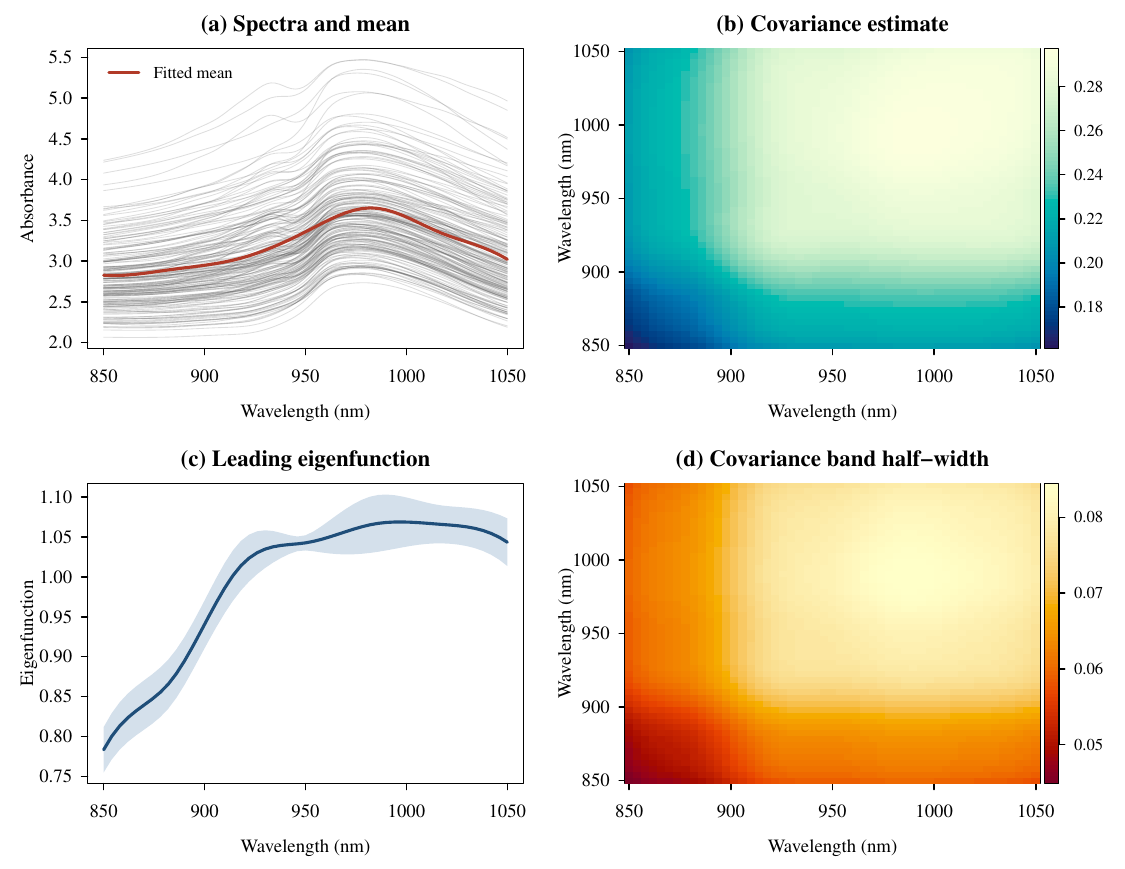}
\caption{Dense Tecator absorbance spectra. (a) Observed spectra and fitted
mean. (b) Estimated covariance. (c) Leading eigenfunction and its
simultaneous band. (d) Half-width of the covariance band. Covariance and
band half-width are on the squared absorbance scale.}
\label{fig:application-tecator}
\end{figure}

A single broad mode accounts for \TecatorFirstSpectrumPercent\% of the
fitted positive spectrum (Figure~\ref{fig:application-tecator}). The
covariance lower band is positive throughout the evaluation grid, and the
leading eigenfunction has the same sign across the wavelength range.
These features describe coordinated variation in absorbance. The eigenfunction's
largest estimated absolute loading occurs at \TecatorPeakWavelength{}~nm.
The corresponding eigenvalue is \TecatorFirstEigenvalue{}, with marginal
interval $[\TecatorFirstEigenvalueLower{},\TecatorFirstEigenvalueUpper{}]$.

To relate this mode to composition, we linearly interpolate the estimated
eigenfunction to the 100 observed wavelengths and compute scores as the
equal-weight inner product with each spectrum after subtracting the
sample mean at each wavelength. {As a result, higher} scores are associated with greater
fat content and lower moisture and protein content. The correlations are
\TecatorFatCorrelation{}, \TecatorMoistureCorrelation{} and
\TecatorProteinCorrelation{}, respectively.
Using four to six internal knots leaves the first eigenvalue and its
positive-spectrum share nearly unchanged. All three correlations agree to three decimal places across these fits.
The loading maximum varies between 998 and 1010~nm. Table~\ref{tab:application-sensitivity} gives the knot comparisons for both datasets.

\begin{table}[htbp]
\centering
\caption{Application estimates with neighboring numbers of internal knots. Peak locations are in years for PBC and nm for Tecator.}
\label{tab:application-sensitivity}
\begin{tabular}{lrrrr}
\toprule
Application & Internal knots & First eigenvalue & Share (\%) & Peak \\
\midrule
PBC & 3 & 0.9378 & 67.3 & 8 \\
PBC & 4 & 0.9725 & 57.8 & 8 \\
PBC & 5 & 0.9786 & 55.7 & 8 \\
Tecator & 4 & 0.2567 & 98.7 & 1010 \\
Tecator & 5 & 0.2566 & 98.7 & 998 \\
Tecator & 6 & 0.2566 & 98.7 & 998 \\
\bottomrule
\end{tabular}
\end{table}
\section{Concluding remarks}\label{sec:conclusion}

We develop a unified framework for simultaneous inference on
functional covariance and eigensystems across sparse-to-dense
sampling designs. By decomposing the residualized subject-level
scores according to pair-overlap patterns, we establish Gaussian
approximations that account for the changing variance contributions
across sampling regimes. These results extend to fixed-index
eigenvalue and eigenfunction inference, as well as joint inference
over a diverging number of eigensystem components.  
The tensor-product B-spline representation facilitates efficient
spectral computation and the construction of confidence sets for
covariance and related eigensystems.

An important direction for future research is to extend the
proposed framework to dependent functional time series. For locally stationary
functional processes, the covariance structure may evolve over
time, raising new challenges for simultaneous inference on
time-varying covariance operators and their eigensystems.
Extensions to dynamical functional principal component analysis,
which accounts for temporal dependence through spectral density
operators, would also be of interest. Such developments would
require Gaussian approximations and spectral perturbation
analysis that accommodate serial dependence and, potentially,
time-varying spectral structures. Another promising direction
is spatial functional data, where dependence across spatial
locations also introduces additional challenges for covariance
estimation and simultaneous spectral inference.

		\section*{Acknowledgments}
 Qirui Hu's research was supported by National Natural Science Foundation of China (NSFC) (Grant Nos.12601520), the Shanghai Engineering Research Center of Finance Intelligence (Grant No.~19DZ2254600) and by TRR 391 \textit{Spatio-temporal Statistics for the Transition of Energy and Transport} (Project number 520388526) funded by the Deutsche Forschungsgemeinschaft (DFG, German Research Foundation).

		\bibliographystyle{apalike}
		\bibliography{ref}
  \clearpage
\end{document}